\documentclass[preprint,11pt]{elsarticle}

\usepackage{amssymb}
\usepackage[utf8x]{inputenc}
\usepackage{amsmath}
\usepackage{mathtools}
\usepackage{amsfonts}
\usepackage{subfigure}
\usepackage{xcolor}

\usepackage[hmargin=2.0cm,vmargin=3.0cm]{geometry}

\newcommand{\GeV}{\,{\rm GeV}}
\newcommand{\MeV}{\,{\rm MeV}}
\newcommand{\tot}{\,{\rm total}}
\newcommand{\baryons}{\,{\rm baryons}}
\newcommand{\mesons}{\,{\rm mesons}}
\newcommand{\PDG}{\,{\rm PDG}}
\newcommand{\cum}{\,{\rm cum}}
\newcommand{\quarks}{\,{\rm quarks}}
\newcommand{\antiquarks}{\,{\rm antiquarks}}
\newcommand{\QCD}{\,{\rm QCD}}

\begin{document}

\begin{frontmatter}

\title{Bag-type Model with Fractal Structure}

\author[a]{Evandro Andrade II}
\author[b,c]{Airton Deppman}
\author[c]{Eugenio Megias}
\author[d]{Débora P. Menezes}
\author[d]{Tiago Nunes da Silva}

\address[a]{Universidade Estadual de Santa Cruz, Ilhéus, Bahia, Brazil}
\address[b]{Universidade de São Paulo, Instituto de Física, São Paulo, Brazil}
\address[c]{Departamento de F{\'\i}sica At\'omica, Molecular y Nuclear and Instituto Carlos I de F{\'\i}sica Te\'orica y Computacional, Universidad de Granada, E-18071 Granada, Spain}
\address[d]{Departamento de Física, CFM, Universidade Federal de Santa Catarina, CP 476, CEP 88.040-900 Florianópolis, SC, Brazil}

\begin{abstract}

In this work we present a bag-type model within a non-extensive statistics applied to the description of the properties of a hadronic system with an underlying fractal structure.
The non-extensive ideal gas inside the bag is determined by the grand canonical partition function from which pressure, energy and particle density as well as temperature and chemical potential 
are obtained for the hadronic system. These quantities are studied in the approximation of fixed mass for all bag constituents but also for discrete and continuum masses. In all cases, the freeze-out 
line, corresponding to the energy per particle equal to 1 GeV and the lines corresponding to a fractal structure inside the proton volume are obtained.
Finally, the pressure on the bag surface of the proton is calculated and the resulting value $(0.135\, {\rm GeV})^4$ obtained.

%In addition, different non-extensive scenarios are investigated. 
\end{abstract}

\begin{keyword}
Bag-type model \sep non-extensive Thermodynamics \sep fractal structure of hadrons

\end{keyword}

\end{frontmatter}

% \linenumbers

\section{Introduction}

The description of the hadron structure is a challenging task because the strong interaction is not completely understood at the non perturbative regime.
Numerical calculations, as Lattice QCD, are haunted, at large values of the chemical potential, by the infamous sign problem and hence, still unable to provide 
non-perturbative results in this regime. This fact allows the investigation of just a few aspects of the hadronic structure from first principles, making phenomenological 
approaches an useful alternative for studying many aspects of hadronic Physics.

Among several phenomenological models of strong interacting systems, Hadron Resonant Gas (HRG) models have been successful in describing several aspects of high energy collisions. 
However, these models have been superseded in recent years by other approaches, such as hydrodynamical models. One of the reasons for the decreasing interest in the original HRG was the 
fact that it has been shown to be useless to explain the distribution of high momentum particles generated in high energy collisions \cite{Hagedorn_HRG}. Recently, though, with the use 
of non extensive statistics, more specifically Tsallis statistics, generalizations of HRG models have proved to be helpful in describing data, evidencing some clear patterns in the 
outcomes of high energy collisions \cite{Cleymans2013}.

In the present work we employ a non extensive HRG model to describe some properties of hadrons. To this end we assume a fractal structure of hadrons, which leads to the emergence of 
non extensivity of the hadronic thermodynamics, which is then described by Tsallis entropic formula. The model considers the hadron as a bag inside which the hadron constituents exert 
a variable pressure. The hadron constituents form an ideal gas, in the non-extensive sense, limited by the hadron volume, and its total energy represents the hadron mass.

The fundamental thermodynamical theory for hadrons was introduced by Hagedorn~\cite{Hagedorn}, Chew and Frautschi~\cite{Geoffrey,Frautschi}. The basic idea in both theories is that 
hadrons are constituted by internal hadrons in a structure which today would be called a fractal~\cite{Mandelbrot}. This structure implies a self-consistency constraint that, added 
to the hypothesis that the compounding particles of hadrons behave as an ideal gas, completely defines the thermodynamical properties of such systems. The unusual hypothesis that an 
ideal gas could be formed by strong interacting particles was demonstrated to be compatible with Schwinger-Dyson
expansion~\cite{DashenMaBernstein}, what triggered a fast development of HRG models~\cite{Venugopalan}.

Hagedorn's  self-consistent theory  gives a hadron mass spectrum formula, as well as the Chew-Frautisch bootstrap model, and in addition predicts a limiting temperature for hadronic matter. 
Latter, based on the MIT bag model \cite{CabibboParisi}, this limiting temperature was assumed to be a phase-transition temperature between the confined and deconfined regimes. Of great 
importance was the fact that Hagedorn's theory predicted the exponential behavior of energy distributions for the outcome of high energy collisions. Although the experimental data available 
at that time confirmed this prediction, with the increase of new data from collisions at higher center of mass energy it became evident that a large tail in the high momentum distributions 
could not be described by the theory,  what caused a decrease in interest on the theory. 

In the last two decades the idea of using Tsallis statistics to extend the distributions obtained from Hagedorn's theory has  gained importance~\cite{Bediaga, Beck, WilkWlodarczyk, Sena, Cleyman_Worku}.
A theory for the non extensive self-consistent thermodynamics was formulated~\cite{Deppman2012, Megias} where the critical temperature is still obtained, but it results in new distributions for energy 
and momentum of the secondary particles in high energy collisions following a power-law distribution that describes correctly the entire range of momentum distributions measured at high energy experiments. 
In addition, a new formula for the hadron mass spectrum is derived from the extended theory, which better describes the spectrum of known hadronic states~\cite{Lucas}. Interestingly, new patterns for 
the limiting temperature and entropic index are obtained, showing that they are nearly independent of the collision energy or the particle species.

Recently, it has been shown that a fractal system like the one proposed by Hagedorn and by Chew and Frautisch must be described by Tsallis statistics instead of the Boltzmann one~\cite{Deppman2016}, 
giving thus a mathematical explanation for the use of Tsallis statistics in the interpretation of high energy data. This fractal structure is also in agreement with the 
information about the fractal dimension for multiparticle production that is obtained from the analysis of intermittency \cite{Deppman2016}. Therefore, the use of Tsallis statistics in HRG models is 
equivalent to the introduction of the fractal structure evidenced in experimental data.

In the present work we investigate the properties of hadrons by introducing a non extensive bag-type model. Our main objective is to investigate how a gas of interacting  fermions and bosons with 
different masses, restricted to move inside a constant volume $V_p=(4 \pi/3)r_o^3$, with $r_o=$1.2~fm, and with total energy fixed to the proton mass, $m_p=$~938~MeV, can give rise to an internal 
pressure equal to the external pressure $B$ used in the well known MIT Bag model. The interaction is included by considering that the particles inside the bag follows a non extensive 
statistics~\cite{Deppman2016, DFMM}. We investigate three different scenarios: a gas with interacting quarks and anti-quarks; a gas of fermions and bosons with different discrete masses; 
a gas of fermions and bosons with masses varying continuously.

\section{Non Extensive Thermodynamics}
\label{sec:non_extensive}

The first step to develop a bag-type model based on a non extensive ideal gas is to find the entropy or the partition function for such gas. Both tasks have already been done elsewhere, and 
they lead to equivalent expressions~\cite{Megias,Pedro}. Firstly, let us define the q-logarithm and the q-exponential functions for particles $(+)$ and anti-particles $(-)$:

\begin{equation}
 \begin{cases}
  \log_{q}^{(+)} (x)=\dfrac{x^{q-1}-1}{q-1}, \hspace{1cm} x \geqslant 0 \\
  \\
  \log_{q}^{(-)} (x)=\dfrac{x^{1-q}-1}{1-q}, \hspace{1cm} x < 0
 \end{cases} \,.
\end{equation}
The q-exponential functions are then given by
\begin{equation}
 \begin{cases}
  e_{q}^{(+)}(x)=\left[1+(q-1)x\right]^{1/(q-1)}, \hspace{1cm} x \geqslant 0 \\
  \\
  e_{q}^{(-)}(x)=\left[1+(1-q)x\right]^{1/(1-q)}, \hspace{1cm} x < 0
 \end{cases} \,.
\end{equation}
The following distribution functions are used to write some of the equations in a more compact form,
\begin{equation}
  \begin{cases}
    n_q^{(+)}(x) = \dfrac{1}{\left(e_q^{(+)}(x) - \xi \right)^q}, \,\,\,\,\,\,\, x \geqslant 0 \\
    n_q^{(-)}(x) = \dfrac{1}{\left(e_q^{(-)}(x) - \xi \right)^{2-q}}, \,\,\,\,\,\,\, x < 0
  \end{cases}  \,.
\label{eqDistFunctions}
\end{equation}

In this study, the properties of the non extensive ideal gas are determined by the grand-canonical partition function
\begin{equation}
\log \Xi(V,T,\mu)_q = -\xi V \int \frac{d^{3}p}{(2\pi)^{3}} \sum_{r=\pm} \Theta(rx) \log_{q}^{(-r)} \left(\frac{e_{q}^{(r)}(x)-\xi}{e_{q}^{(r)}(x)}\right)  \,,
\label{eqPartFunc}
\end{equation}
where $x=\beta(E_p-\sum_a \mu_a Q_a)$, with $E_p = \sqrt{p^2 + m^2}$ is the particle energy, $Q_a$ are conserved charges and $\mu_a$ the chemical potentials associated to these charges. 
$\xi = \pm 1$, for bosons and fermions respectively and $\Theta$ is the step function. In the $(uds)$ flavor sector of QCD the only conserved charges are the electric charge~$Q$, the baryon 
number $B$, and the strangeness~$S$. 

All thermodynamical functions for such a non extensive ideal gas~\cite{Tsallis1988} can be obtained by applying the usual thermodynamics relations, and, for the case of interest in the present
work, the relevant quantities were already obtained elsewhere~\cite{Megias,Pedro}, so here they are only listed.

The thermal expectation value for the charge $Q_a$ is given by
\begin{equation}
\left\langle Q_a \right\rangle = \beta^{-1} \frac{\partial}{\partial \mu_a} \log \Xi_q \big|_{\beta} \,.  \label{eq:Qa1}
\end{equation} 
This can be expressed as well in the form
\begin{equation}
\langle Q_a\rangle = Q_a \langle N \rangle \,, \label{eq:Qa2}
\end{equation}
where $\langle N \rangle$ is the averaged number of particles. In the following we consider only the effects of the baryon number in the thermal medium, so that we switch on only the baryonic 
chemical potential. In the following explicit expressions we use $x = \beta(E_p - \mu_B B)$. According to Eq.~(\ref{eqPartFunc}), and using Eqs.~(\ref{eq:Qa1}) and (\ref{eq:Qa2}), the average 
number of particles is
\begin{equation}
 \left\langle N \right\rangle = V \left[ C_{N,q}(\mu_B,B,\beta,m) + \dfrac{1}{2\pi^2}\displaystyle\int_0^{p_*} dp \, p^2 n_q^{(-)}(x) +  \dfrac{1}{2\pi^2}\displaystyle\int_{p_*}^\infty dp \, p^2 n_q^{(+)}(x) \right] 
 \,,  \label{eq:Npart}
\end{equation}
where $p_* = \sqrt{(\mu_B B)^2 - m^2} \cdot \Theta(\mu_B B - m)$, and 
\begin{equation}
 C_{N,q}(\mu_B,B,\beta,m) = \dfrac{1}{2\pi^2}\dfrac{\mu_B B \sqrt{(\mu_B B)^2 - m^2}}{\beta} \dfrac{2^{q-1}+2^{1-q}-2}{q-1}\Theta(\mu_B B -m) \,.
\end{equation}
Note that the average baryon number can be computed from Eqs.~(\ref{eq:Qa2}) and (\ref{eq:Npart}). In the case that the constituents are hadrons, i.e. mesons, baryons and antibaryons, this leads to
\begin{equation}
    \langle B \rangle = \langle N_{\textrm{baryons}} \rangle  - \langle N_{\textrm{antibaryons}} \rangle \,.
\end{equation}

The average total energy is given by
\begin{equation}
 \left\langle E \right\rangle = -\dfrac{\partial}{\partial \beta} \log \Xi_q \bigg|_{\mu_B} + \dfrac{\mu_B}{\beta} \dfrac{\partial}{\partial \mu_B} \log \Xi_q \bigg|_{\beta},
\end{equation}
resulting in
\begin{equation}
 \left\langle E \right\rangle = V \left[ C_{E,q}(\mu_B,B,\beta,m) + \dfrac{1}{2\pi^2}\displaystyle\int_0^{p_*} dp \, p^2 \, E_p  n_q^{(-)}(x)  + \dfrac{1}{2\pi^2}\displaystyle\int_{p_*}^\infty dp \,
 p^2  \, E_p  n_q^{(+)}(x)  \right] \,,
\end{equation}
with 
\begin{equation}
C_{E,q}(\mu_B,B,\beta,m) = \mu_B B \cdot C_{N,q}(\mu_B,B,\beta,m) \,.
\end{equation}
The pressure is given by
\begin{equation}
P = \dfrac{1}{\beta}\dfrac{\partial}{\partial V}\log\Xi_q \,,
\end{equation}
which leads to
\begin{equation}
 \left\langle P \right\rangle =  -\xi \dfrac{T}{2\pi^2} \left[ \displaystyle\int_0^{p_*} dp \, p^2 \, \text{log}_q^{(+)} \left( \dfrac{1}{1 + \xi n^{(-)}_q(x)^{1/(2-q)}} \right) + 
 \displaystyle\int_{p_*}^\infty dp \, p^2 \, \text{log}_q^{(-)} \left( \dfrac{1}{1 + \xi n^{(+)}_q(x)^{1/q}} \right) \right] \,.
\end{equation}
Finally, the entropy is
\begin{equation}
S = -\beta^2 \frac{\partial}{\partial\beta} \left( \frac{\log\Xi_q}{\beta} \right) \bigg|_{\mu_B} \,,
\end{equation}
and this can be explicitly expressed as
\begin{eqnarray}
\langle S \rangle &=& V   \bigg[ \displaystyle\int_0^{p_*} dp \, p^2 \,  \Big\{ - n_q^{(-)}(x) \log_q^{(+)}(n_q^{(-)}(x)^{1/(2-q)})  + \xi(1+\xi n_q^{(-)}(x)^{1/(2-q)})^{(2-q)}  \times  \nonumber \\
&&\qquad\qquad\qquad \times \, \text{log}_q^{(+)}\left( 1 + \xi n^{(-)}_q(x)^{1/(2-q)} \right) \Big\}  \\
&&\quad + \displaystyle\int_{p_*}^\infty dp \, p^2 \, \Big\{ - n_q^{(+)}(x) \log_q^{(-)}(n_q^{(+)}(x)^{1/q})  + \xi(1+\xi n_q^{(+)}(x)^{1/q})^q \text{log}_q^{(-)} \left( 1 + \xi n^{(+)}_q(x)^{1/q} \right) \Big\}  \bigg] \,.  \nonumber 
\end{eqnarray}

\section{Fractal Bag Model at different scenarios - formalism and results}

For building the fractal bag-type model, we assume that the hadrons can be described as a bag where a non extensive ideal gas at temperature $T$ exerts a pressure on the bag membrane that confines it. 
By describing the gas as a non extensive gas, interactions between the compound particles are automatically included. Such interaction is supposed to form a fractal structure that needs to be described 
by Tsallis statistics as shown in ~\cite{Deppman2016} and discussed in the Introduction. In the following we study the gas in three different scenarios: fixed mass, where all particles have the same mass, 
chosen to be the current quark mass; the discrete mass, where constituent masses can be different, but are constrained to take the value of one of the observed hadron masses; continuous mass where 
constituent masses can vary continuously according to the generalized Hagedorn mass spectrum.

It is interesting to evaluate the temperature at which the total energy of the system is equal to the proton mass, $m_p=938$~MeV. The total energy is obtained by multiplying the proton volume by the 
energy density, with $V_p=(4 \pi/3)r_o^3$, with $r_o=$1.2~fm. We also define the freeze-out line by imposing that $\varepsilon/n = 1\GeV$. This definition follows from the calculations performed in 
\cite{Megias}, where it was shown that this specific curve can explain heavy ion collision experimental results if $q=1.14$. This picture of the proton as a bag will be investigated in all three 
aforementioned mass scenarios. 

One important aspect to study are the limits of temperature and chemical potential within which the proton is supposed to exist as a confined system, according to our model. In a $T$ vs $\mu_B$ diagram, 
the region below the freeze-out line refers to a confined regime whilst the entire region above the freeze-out line refers to a deconfined regime.

Other aspects of this fractal bag model are also examined in the following sections.

\subsection{Fixed mass}

The fractal structure we implement when using Tsallis statistics imposes that the bag is formed by partons which have also internal structure which are themselves partons. This is similar to the 
fireball by Hagedorn, or to the bootstrap principle by Chew and Frautschi. Therefore, this first scenario is studied as an exercise since it is based only on quarks of fixed mass. And since quarks 
themselves don't have internal structure, the self-similarity present in both Hagedorn and Chew and Frautschi models is not fulfilled here. 

As a first step in our calculations we consider that the all particles within the gas have a single mass, $m$, which is fixed to a light quark mass, that
is, $m=2.3$~MeV. We consider for the degeneracy of quarks
\begin{equation}
g_{\quarks} = 2 N_f \,, 
\end{equation}
with $N_f = 3$, while the factor 2 comes from the spin. We don't
include a factor $N_c$ in the degeneracy, as it is assumed that the
quarks form color singlet states. The degeneracy of anti-quarks is the
same, i.e. $g_{\antiquarks} = g_{\quarks}$. Finally, note that the
baryon number of quarks and anti-quarks is $B_{\quarks} = 1/3$ and
$B_{\antiquarks} = -1/3$. The average baryon number is computed using the formalism of Sec.~\ref{sec:non_extensive}, and the expression reads
\begin{equation}
\langle B \rangle =  \frac{1}{3} \left( \langle N_{\textrm{quarks}} \rangle  - \langle N_{\textrm{antiquarks}} \rangle  \right) \,.
\end{equation}
The energy density of such system depends both on the temperature $T$ and on the baryonic chemical potential,
$\mu_B$.

\begin{figure*}[ht]
  \begin{center}
    \subfigure[]{
     \label{fig:Tmu_FixedMass}
     \includegraphics[width=0.45\textwidth]{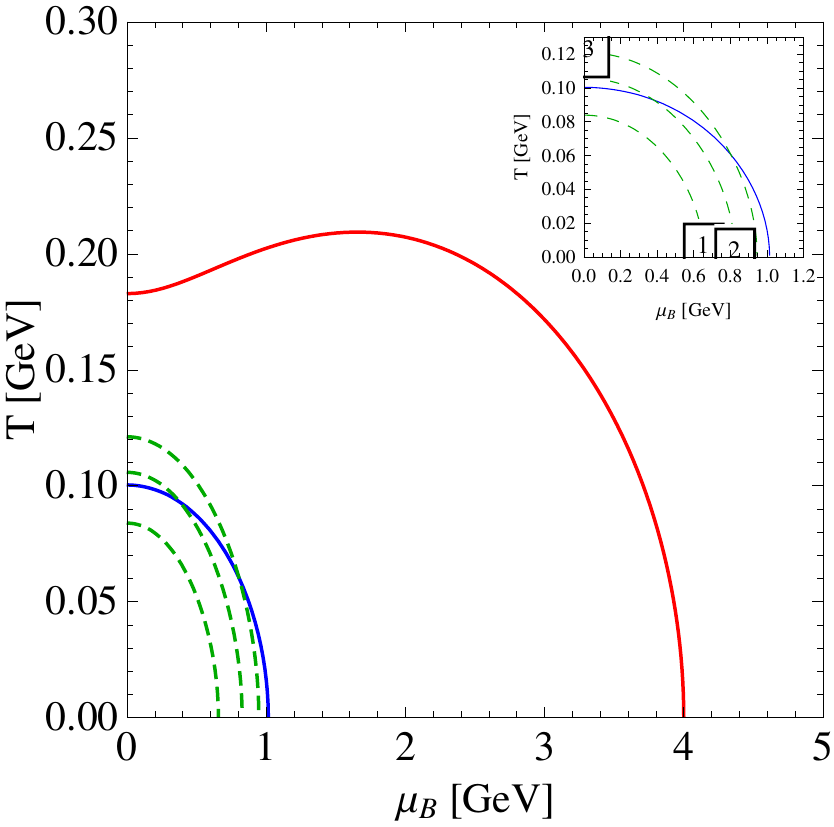}}
     \subfigure[]{
     \label{fig:N_FixedMass}
     \includegraphics[width=0.45\textwidth]{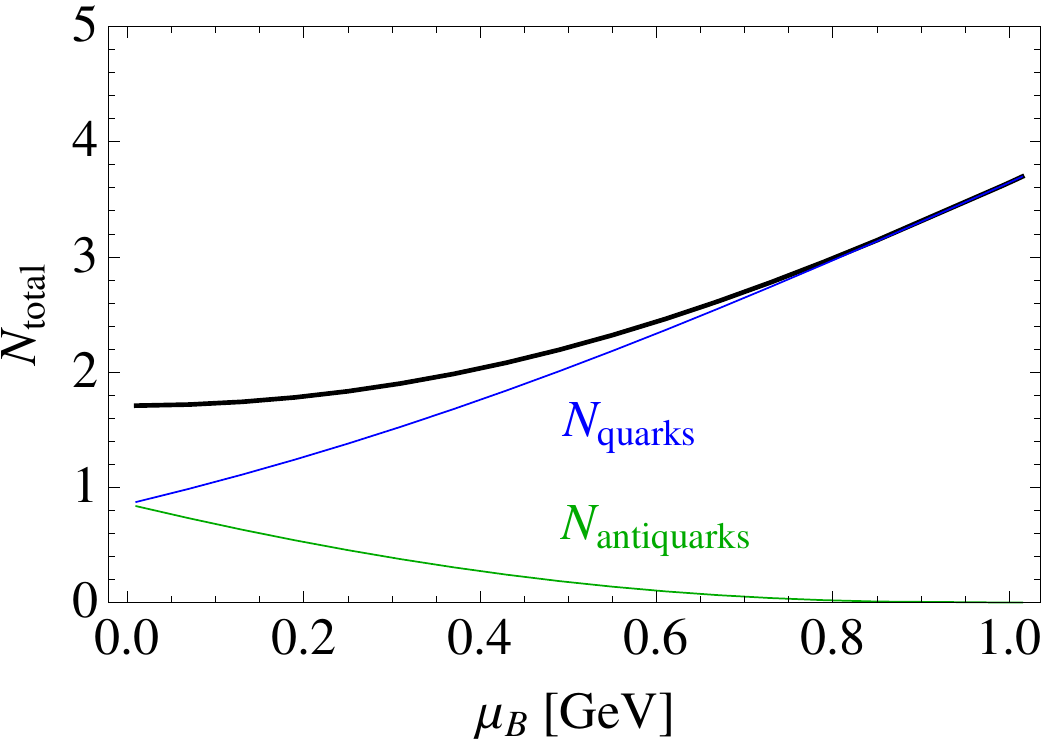}
     } 
     \caption{(a) Temperature as a function of baryonic chemical potential for a non-extensive gas with total energy inside a volume~$V_p$ equal to the proton mass. We display as solid red
       line the freeze-out line $\varepsilon/n = 1\GeV$. The dashed green lines correspond to the region in which the total number of particles inside a volume $V_p$ are equal to 1, 2 and 3.
       The inserted figure is a zoom of the of the main figure. (b) Number of quarks and of anti-quarks (inside a volume $V_p$) as a function of the baryonic chemical potential, along the line
       of $\varepsilon \cdot V_p = m_p$ (blue line in Fig.~\ref{fig:Tmu_FixedMass}.) The solid black line is $N_{\tot} = N_{\quarks} + N_{\antiquarks}$.}
     \label{fig:FixedMass}
  \end{center}   
\end{figure*}

In Fig.~\ref{fig:Tmu_FixedMass} we present different situations: the line corresponding to the total energy of the system defined as mentioned above, the freeze-out line and the lines delimiting 
the region in which the total number of particles inside a volume $V_p$ are equal to 1, 2 and 3. In all cases, the temperature decreases as the chemical potential increases, what results in 
larger masses at lower temperatures. As the chemical potential vanishes, the temperature increases up to $T \sim 180$~MeV. This result, although qualitatively correct, causes some problems in the 
context of the non extensive model of hadrons. According to Ref. \cite{Megias}, we would expect that the red line would cross the blue line, but this is not the case here. Indeed, at temperatures 
above $T \sim 61$~MeV the system is known to have crossed the phase transition point, therefore no hadrons should exist at such temperature, as can be seen in Ref.~\cite{Lucas}.
The inconsistency found here is due, as we will show in the following, to the fact that we are using a fixed mass for the constituent 
particles. 

In Fig.~\ref{fig:N_FixedMass}, we show the
number of quarks and of anti-quarks inside the volume $V_p$ as a function of the baryonic chemical potential. We can see that the total number of particles increases with the chemical potential, as expected.

\subsection{Discrete masses}

In the discrete mass scenario the self-similarity can be understood in terms of quarks and gluons. If one consider the hadron as a bag composed of a sea of quarks and gluons then one should 
always find pairs or triplets of quarks correlated in such a way as to form structures similar to the hadrons we observe in nature, similar in the sense that they present the same quantum numbers.

With this picture in mind, we consider the case of different masses for those particles inside the bag. This is done by summing the energy and pressure for all
different masses according to the masses and multiplicity of the known hadrons. Apart from the summation, all the formulae for the thermodynamical quantities are the same. 

The second step is to consider that, according to Hagedorn's and Chew-Frautisch models, the 
compound particles of hadrons have a variety of masses, and we include such masses by considering that the gas inside the
bag can be formed by particles with masses that correspond to the masses of known hadrons, and considering their multiplicity. 
By dividing the thermodynamical relations mentioned in the previous section by the volume and summing over all hadrons in the bag, 
we get the particle density of hadrons to be

\begin{equation}
 n = \frac{\langle N \rangle}{V} = \sum_i g_i \left[  C_{N,q}(\mu_B,B_i,\beta,m_i)  +  \frac{1}{2\pi^2} \int_0^{p_{i*}} dp\, p^2  \, n_q^{(-)}(x_i) +  \frac{1}{2\pi^2}  \int_{p_{i*}}^\infty dp 
 \, p^2 \, n_q^{(+)}(x_i)  \right] \,,
     \label{eqBarDens}
\end{equation}
where $x_i = \beta(E_i - \mu_B B_i)$ and $p_{i*} = \sqrt{(\mu_B B_i)^2 - m_i^2} \cdot \Theta(\mu_B B_i - m_i)$. $E_i = \sqrt{p^2 + m_i^2}$ is the energy, $g_i$~is the degeneracy and $B_i$ the baryon number 
of the particle $i$. Notice that $B_i = +1, -1, 0$ for baryons, antibaryons and mesons, respectively.~\footnote{Note that, when assuming $\mu_B \ge 0$, then $p_{i*}$ can have a nonvanishing value only for baryons. 
$p_{i*} = 0$ for anti-baryons $(B_i = -1)$ and mesons $(B_i = 0)$, so that in these cases the only relevant integrals are those involving $n_q^{(+)}(x)$.} The baryon density can be computed from 
Eq.~(\ref{eqBarDens}), leading to

\begin{equation}
    \langle n_B \rangle = \sum_{i} B_i \frac{\langle N_i \rangle}{V} = \langle n_{\textrm{baryons}} \rangle - \langle n_{\textrm{antibaryons}} \rangle \,.
\end{equation}

 The number of (anti)baryons is obtained from Eq.~(\ref{eqBarDens}) by restricting the sum over the spectrum of (anti)baryons only, and considering the corresponding quantum numbers. Finally, the 
 energy density is given by

\begin{equation}
 \varepsilon = \frac{\langle E \rangle}{V} = \sum_i g_i \left[  C_{E,q}(\mu_B,B_i,\beta,m_i)  +  \frac{1}{2\pi^2} \int_0^{p_{i*}} dp\, p^2  \, E_i n_q^{(-)}(x_i) +  \frac{1}{2\pi^2} 
 \int_{p_{i*}}^\infty dp \, p^2 \, E_i n_q^{(+)}(x_i)  \right] \,,
     \label{eqEnDens}
\end{equation}

and the pressure is calculated according to

\begin{equation}
 P =  -\frac{T}{2\pi^2} \sum_i g_i \xi_i \left[ \displaystyle\int_0^{p_{i*}} dp \, p^2 \,  \text{log}_q^{(+)} \left( \dfrac{1}{1 + \xi_i n^{(-)}_q(x_i)^{1/(2-q)}} \right) +  
 \displaystyle\int_{p_{i*}}^\infty dp \, p^2 \, \text{log}_q^{(-)} \left( \dfrac{1}{1 + \xi_i n^{(+)}_q(x_i)^{1/q}} \right)  \right] \,,
 \label{eqBarP}
\end{equation}

with $\xi_i = -1$ for (anti)-baryons, and $\xi_i = 1$ for mesons.

\begin{figure}[!ht]
  \begin{center}
     \includegraphics[width=0.45\textwidth]{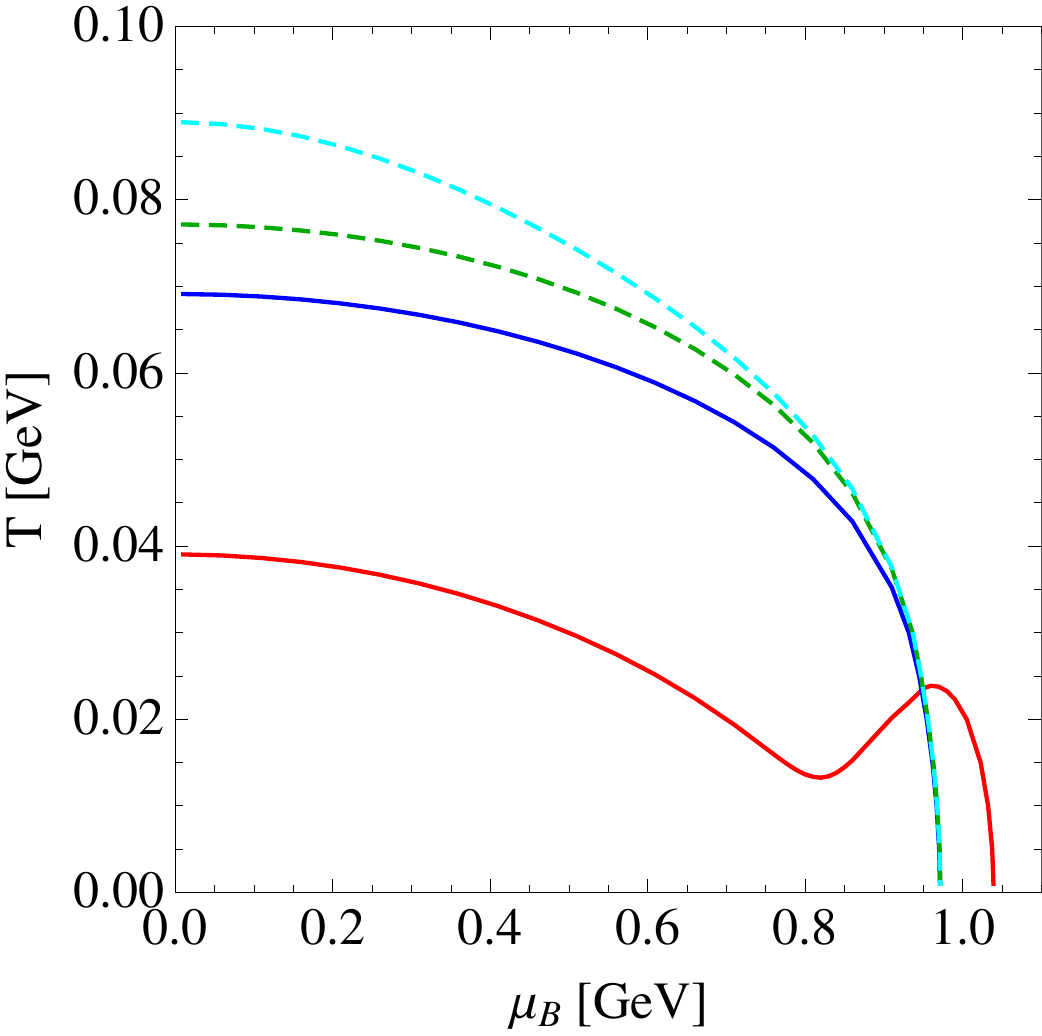}
     \caption{Temperature as a function of baryonic chemical potential for a non-extensive gas with total energy inside a volume~$V_p$ equal to the proton mass (solid blue line). We display as 
     solid red line the freeze-out line $\varepsilon/n = 1\GeV$. The dashed lines correspond to the region in which the total number of particles (dashed green line) and the total number of baryons (dashed
       cyan line) inside a volume $V_p$ are equal to 1, respectively. These results have been obtained for the case of discrete masses and $q=1.14$.}
     \label{figTmu_discretemasses}
  \end{center}   
\end{figure}

\begin{figure}[!ht]
  \begin{center}
     \includegraphics[width=0.45\textwidth]{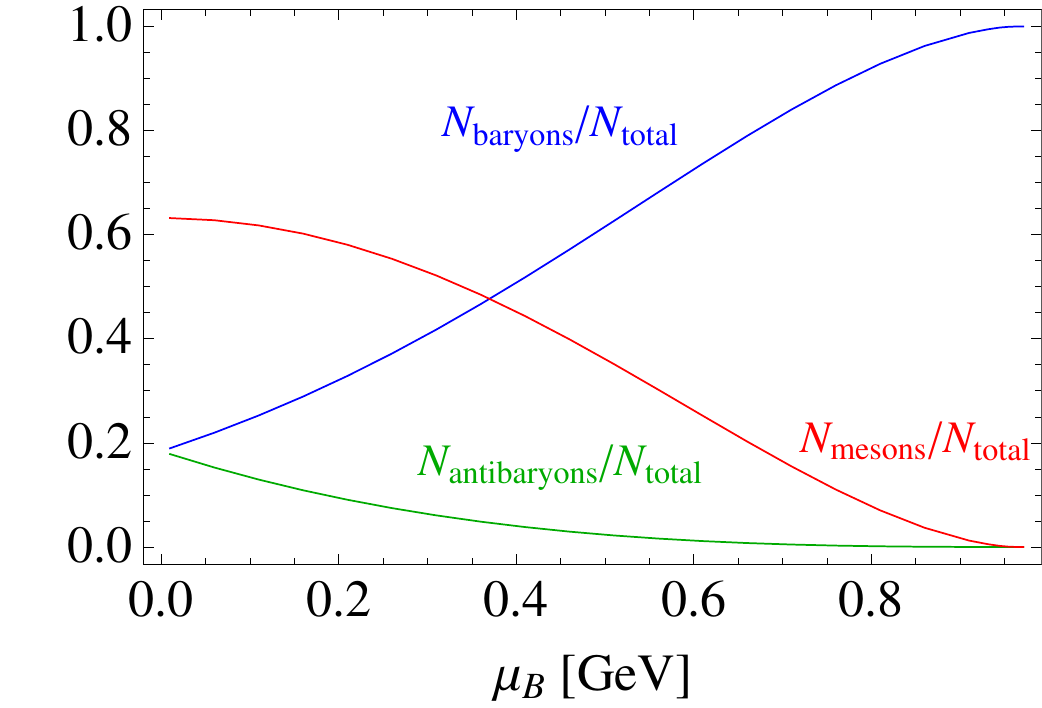} 
     \caption{Number of baryons, antibaryons and mesons (inside a volume $V_p$) as a function of the baryonic chemical potential, along the line of $\varepsilon \cdot V_p = m_p$ (blue line 
     in Fig.~\ref{figTmu_discretemasses}.)  These quantities are normalized to $N_{\tot} = N_{\baryons} + N_{\mesons}$. The solid black line in the right panel is $N_{\tot}$.  
    See Fig.~\ref{figTmu_discretemasses} for further details.}
     \label{fig:Nbm}
  \end{center}   
\end{figure}

In Figure \ref{figTmu_discretemasses} we show the $T$ vs $\mu_B$ diagram for the case of discrete masses. The critical line, for which $\epsilon/n = 1$ GeV is represented by a solid red line while 
the solid blue line represents a gas with total energy equal to the proton mass. Since a given system is confined only in the region below the critical line, we can see that the proton can be found 
inside a narrow range of baryonic chemical potential, around $\mu_B = 0.96$ GeV and for temperatures not higher than $0.025$ GeV.

Figure \ref{fig:Nbm} shows the number of baryons, antibaryons and mesons normalized to the total number of particles. As expected, the number of baryons increase with increasing baryonic chemical
potential as the amount of antibaryons and mesons decrease. Also, according to Figure \ref{figTmu_discretemasses}, the proton exists close to $\mu_B = 0.96$ GeV. At this value of chemical potential,
Figure \ref{fig:Nbm} shows that the proton is completely baryonic in content. This result exemplifies the consistency of the model.

\subsection{Continuous mass}

The third step is to consider a continuous mass spectrum in the computation. It has been argued in some previous works that the hadron mass spectrum can be understood in a self-consistent way by 
using non-extensive statistics. In particular, the spectrum density is given by~\cite{Lucas}
\begin{equation}
\rho(m) = \gamma m^{-5/2} e_q^{(+)}(m/T_o) \,, \label{eq:rhom1}
\end{equation}
where $\gamma$ and $T_o$ are parameters to be determined that characterize the spectrum. In the present work we are interested in the thermodynamics at finite temperature and baryonic chemical 
potential, so that it is important to distinguish between mesons and baryons, and we will use a density for mesons, $\rho_{\rm mesons}(m)$, and a density for baryons, $\rho_{\rm baryons}(m)$. 

Using these densities, we find for the thermodynamic quantities expressions similar to Eqs.~(\ref{eqBarDens})-(\ref{eqBarP}), but replacing the discrete summation in states by a continuous 
integral in the mass, i.e.
\begin{equation}
\sum_i g_i \longrightarrow  \int_0^\Lambda  dm \sum_{\xi=-1,1} \rho_\xi(m) \,.
\end{equation}
In this formula we are using the notation $\rho_{1}(m) = \rho_{\rm mesons}(m)$ for the density of mesons, and $\rho_{-1}(m) = \rho_{\rm baryons}(m)$ for the density of baryons. In addition, 
we will consider $\Lambda = 2.0 \, \textrm{GeV}$ as a cut-off in the spectrum. Then, the thermodynamic quantities can be computed with the formulas
\begin{equation}
 n := \frac{\langle N \rangle}{V} =  \int_{0}^\Lambda dm \sum_{\xi=-1,1} \rho_\xi(m)  \left[  C_{N,q}(\mu_B,B_\xi,\beta,m)  +  \frac{1}{2\pi^2} \int_0^{p_{*}} dp\, p^2  \, n_q^{(-)}(x) +  
 \frac{1}{2\pi^2}  \int_{p_{*}}^\infty dp \, p^2 \, n_q^{(+)}(x)  \right] \,,
     \label{eqBarDenscontinuous}
\end{equation}
for the particle density of hadrons,
\begin{equation}
 \varepsilon := \frac{\langle E \rangle}{V} = \int_{0}^\Lambda dm \sum_{\xi=-1,1} \rho_\xi(m) \left[  C_{E,q}(\mu_B,B_\xi,\beta,m)  +  \frac{1}{2\pi^2} \int_0^{p_{*}} dp\, p^2  \, E 
 \, n_q^{(-)}(x) +  \frac{1}{2\pi^2}  \int_{p_{*}}^\infty dp \, p^2 \, E\,  n_q^{(+)}(x)  \right] \,,
     \label{eqBarEDenscontinuous}
\end{equation}
for the energy density, and 
\begin{eqnarray}
P &=&  -\frac{T}{2\pi^2} \int_{0}^\Lambda dm \sum_{\xi=-1,1} \xi \cdot \rho_\xi(m) \Bigg[ \displaystyle\int_0^{p_{*}} dp \, p^2 \,  \text{log}_q^{(+)} \left( \dfrac{1}{1 + \xi n^{(-)}_q(x)^{1/(2-q)}} \right) \nonumber \\
&&\qquad\qquad\qquad\qquad\qquad\qquad\qquad\qquad\qquad\qquad +  \displaystyle\int_{p_{*}}^\infty dp \, p^2 \, \text{log}_q^{(-)} \left( \dfrac{1}{1 + \xi n^{(+)}_q(x)^{1/q}} \right)  \Bigg] \,, \label{eqBarPcontinuous}
\end{eqnarray}
for the pressure. Note that in these formulas
\begin{equation}
E = \sqrt{p^2 + m^2}, \hspace{1cm} {x = \beta(E - \mu_B B_\xi)}, \hspace{0.5cm} \text{and} \hspace{0.5cm} p_{*} = \sqrt{(\mu_B B_\xi)^2 - m^2} \cdot \Theta(\mu_B B_\xi - m)
\end{equation} 
are functions of the continuous parameter $m$. In addition, while the baryon number of mesons is zero, i.e. $B_1 = 0$, half of the baryons have baryon number $B_{-1} =  1$, and the other 
half have baryon number $B_{-1} = -1$, i.e. they can be particles or antiparticles. This means that in the previous formulas the density of baryons should be understood as
\begin{equation}
\rho_{-1}(m) \to \frac{1}{2} \rho_{-1}(m) \delta_{B_{-1}=1} +  \frac{1}{2} \rho_{-1}(m)  \delta_{B_{-1}=-1} \,.
\end{equation}
Then, as mentioned above for the discrete mass spectrum, it is obvious that $p_{*}$ can have a non-vanishing value only for baryons $(B_{-1}=1)$.

\begin{figure}[!ht]
  \begin{center}
     \subfigure[]{
     \label{fig:Ncum_mesons}
     \includegraphics[width=0.45\textwidth]{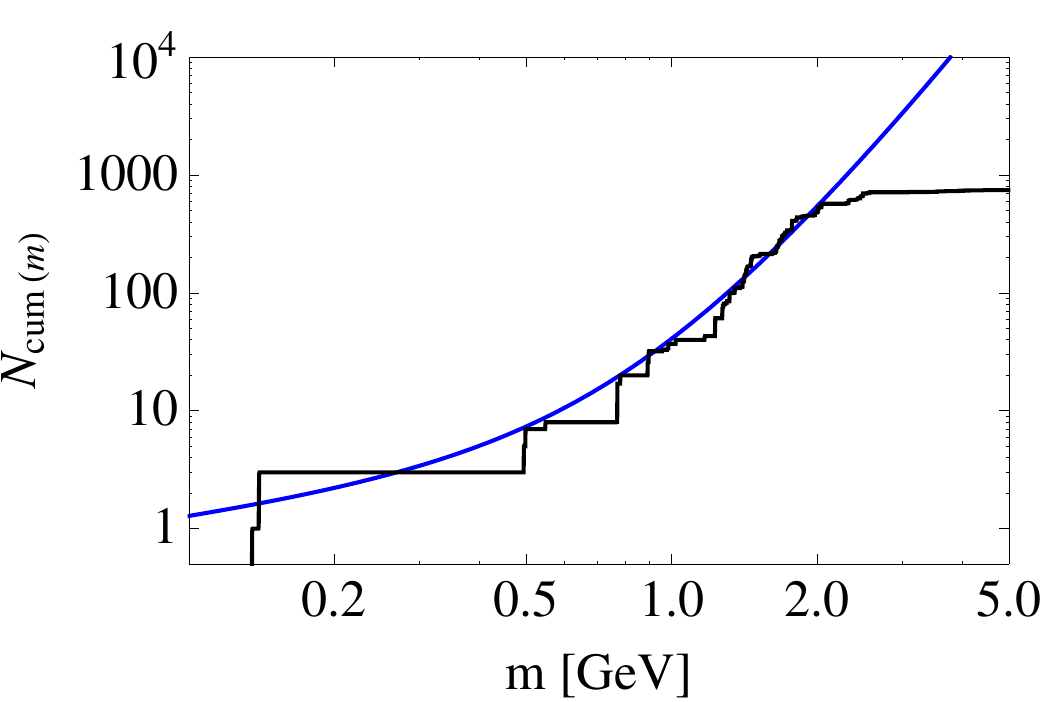} 
     }
     \subfigure[]{
     \label{fig:Ncum_baryons}
     \includegraphics[width=0.45\textwidth]{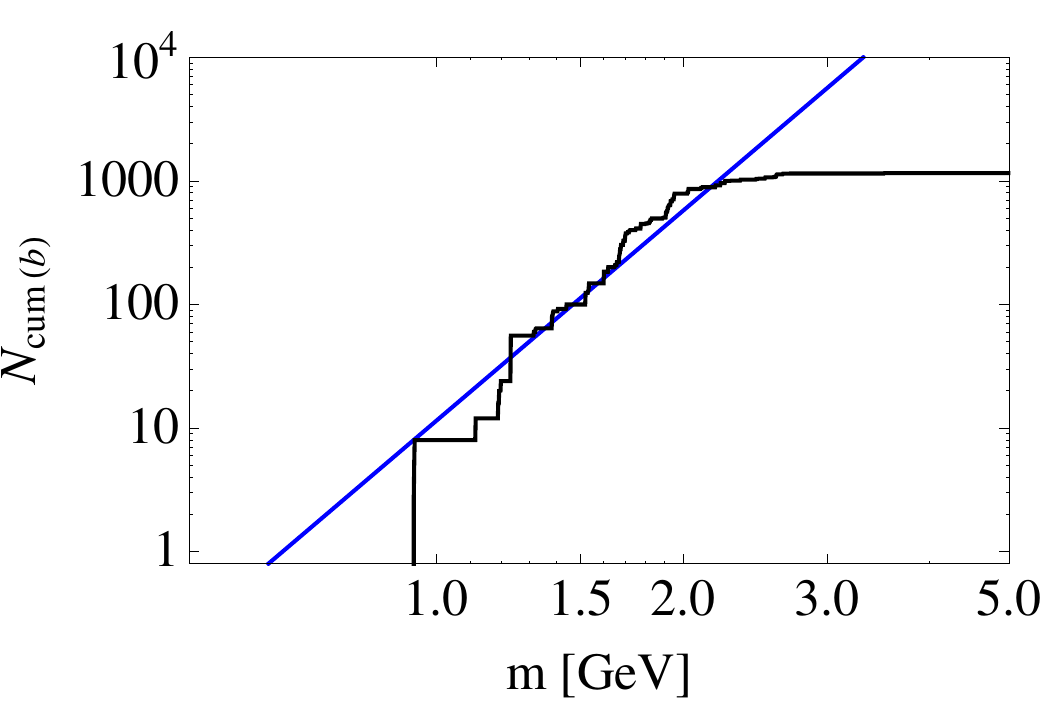}
     } 
     \caption{Cumulative number of meson spectrum, Fig.~\ref{fig:Ncum_mesons}, and baryon spectrum, Fig.~\ref{fig:Ncum_baryons}, from the PDG (continuous black lines), and from the q-exponential 
     fit~(continuous blue lines). The fits are performed by using Eq.~(\ref{eq:N}), and the values of the parameters $(\gamma,T_o)$ are reported in Eq.~(\ref{eq:param_mesons}) for mesons, and 
     Eq.~(\ref{eq:c}) for baryons. In the latter case the distribution turns out to be the one given by Eq.~(\ref{eq:dist_baryons}).}
     \label{fig:Ncum}
  \end{center}   
\end{figure}

We show in Fig.~\ref{fig:Ncum} the cumulative number of mesons and baryons. It is defined as the number of hadrons below some mass $m$, i.e.
\begin{equation}
N_{\cum}(m) = \sum_i g_i \Theta(m-m_i) \,,
\end{equation}
where $g_i$ is the degeneracy factor, and $m_i$ is the mass of the $i$-th hadron. The density of states is given by $\rho(m) = dN_{\cum}(m)/dm$. According to the non-extensive self-consistent theory, 

In order to compute the cumulative number, it is convenient to add in Eq.~(\ref{eq:rhom1}) some contribution so that the integral $\int_\epsilon^m d{\tilde m} \, \rho({\tilde m})$ is finite in the 
limit $\epsilon \to 0$. This contribution is $\rho(m) \to \rho(m) - \gamma m^{-5/2}(1 + m/T_o)$, and it only affects appreciably the low $m$ behavior of $\rho(m)$, in particular for masses smaller 
than the pion mass. Then the cumulative number is
\begin{equation}
N_{\cum}(m) = \int_0^m d{\tilde m} \, \rho({\tilde m}) = \frac{2\gamma}{3 m^{3/2}} \left( - \, {}_2F_1\left[-\frac{3}{2}, -\frac{1}{q-1}, -\frac{1}{2}, -(q-1)\frac{m}{T_o} \right]  + 1 + \frac{3m}{T_o} \right) \,, \label{eq:N}
\end{equation}
where ${}_2F_1$ is the ordinary hypergeometric function. As we are studing the phase diagram in terms of the baryonic chemical potential, it is convenient to distinguish between meson and hadron spectrum. 
The best fit of the meson spectrum from the PDG with Eq.~(\ref{eq:N}) leads to the following values
\begin{equation}
\gamma = 10.36 \cdot 10^{-3} \,, \qquad T_o = 51.05 \MeV \,, \label{eq:param_mesons}
\end{equation}
and it is displayed in Fig.~\ref{fig:Ncum_mesons}. In this fit we have imposed that the theoretical distribution of Eq.~(\ref{eq:N}) reproduces very well the low $m$ region, i.e. the pion mass regime, 
as this is the most important contribution in the partition function.~\footnote{This can be done by assumming a very small error bar for the experimental distribution $N_{\PDG}(m)$ in the regime 
$m \sim m_\pi$, when computing the $\chi^2/$dof and its minimization.} In the case of baryons, the best fit to the experimental distribution of the PDG corresponds to values\begin{equation}
T_o \to 0 \,, \qquad \gamma \to 0 \,, \qquad \textrm{so that} \qquad \frac{T_o}{\gamma^{q-1}} = c \,, \label{eq:fitbaryons}
\end{equation}
with $c = 0.078 \GeV$, and it is displayed in Fig.~\ref{fig:Ncum_baryons}. For the same reason as in the meson case, we have imposed that the theoretical distribution reproduces very well the regime 
of the nucleon mass. Then, the result of Eq.~(\ref{eq:fitbaryons}) can be easily obtained by solving the equation~$N(m_n) = g_n$ with $N(m)$ given by Eq.~(\ref{eq:N}) and $g_n = 8$, and taking the 
limit $T_o \to 0$ and $\gamma  \to 0$. This leads to
\begin{equation}
\frac{T_o}{\gamma^{q-1}} = c \qquad \textrm{with} \qquad c = (q-1) \left(\frac{2(q-1)}{(5-3q)g_n m_n^{ 3/2}} \right)^{(q-1)} \cdot m_n  \simeq 0.078  \GeV\,, \label{eq:c}
\end{equation}
when $q=1.14$ and $m_n = 0.938 \GeV$. Finally, note that in the limit of Eq.~(\ref{eq:fitbaryons}) the baryon-mass density and cumulative number behave as
\begin{equation}
\rho_b(m) \simeq \left( \frac{q-1}{c} \right)^{1/(q-1)} m^{\frac{(7-5q)}{2(q-1)}} \,, \qquad N_{\cum (b)}(m) \simeq 2 \left(\frac{q-1}{5-3q} \right) \left( \frac{q-1}{c} \right)^{1/(q-1)} m^{\frac{(5-3q)}{2(q-1)}} \,, \label{eq:dist_baryons}
\end{equation}
and $N_{\cum (b)}(m_n) = g_n$ is consistent with the expression for $c$ given by Eq.~(\ref{eq:c}). Notice that the distributions in Eq.~(\ref{eq:dist_baryons}) depend on just one parameter. 
It is remarkable that the experimental baryon distribution from the PDG can be reproduced so well by fitting just one single parameter.

\begin{figure}[!ht]
  \begin{center}
     \includegraphics[width=0.35\textwidth]{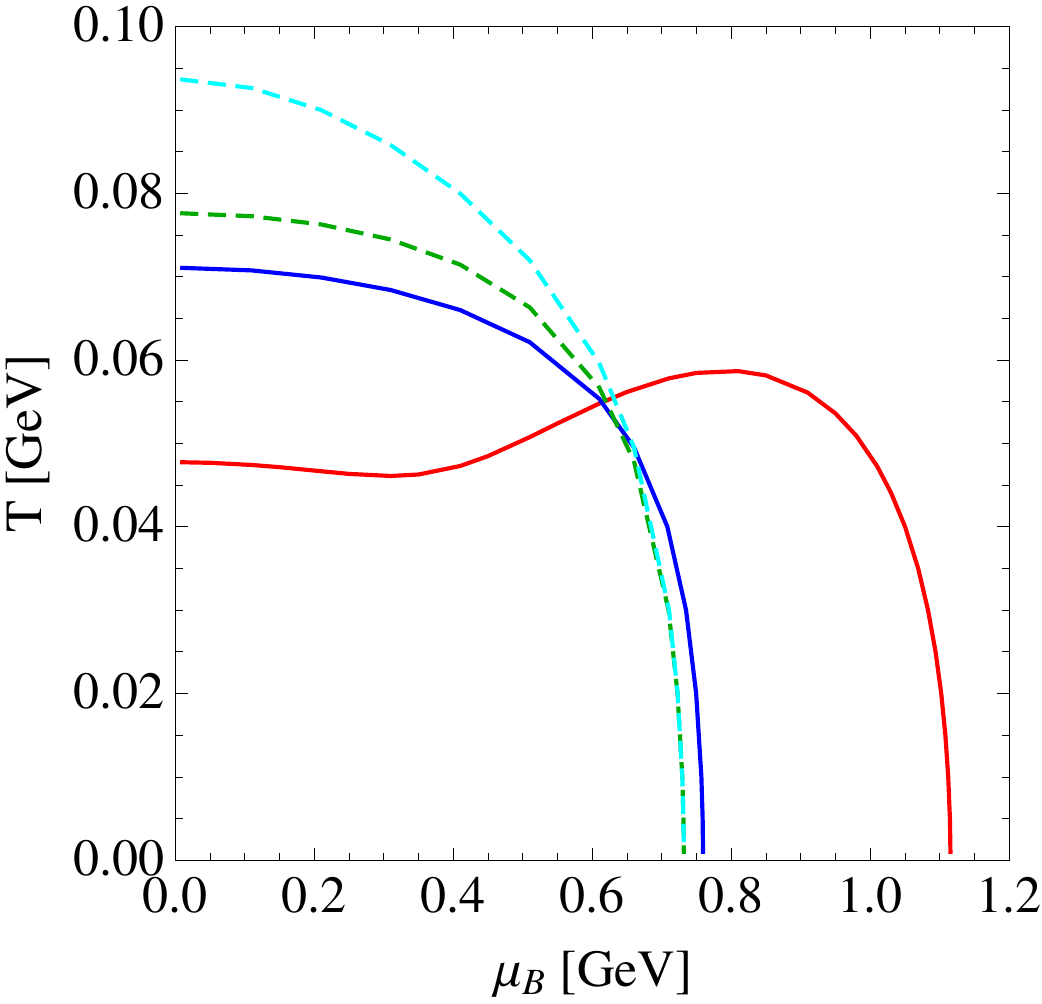}
     \caption{Same plot as Fig.~\ref{figTmu_discretemasses}, for the continuous mass spectrum. See caption of Fig.~\ref{figTmu_discretemasses} for further details.}
     \label{figTmu_continuous}
  \end{center}   
\end{figure}

\begin{figure}[!ht]
  \begin{center}
     \includegraphics[width=0.35\textwidth]{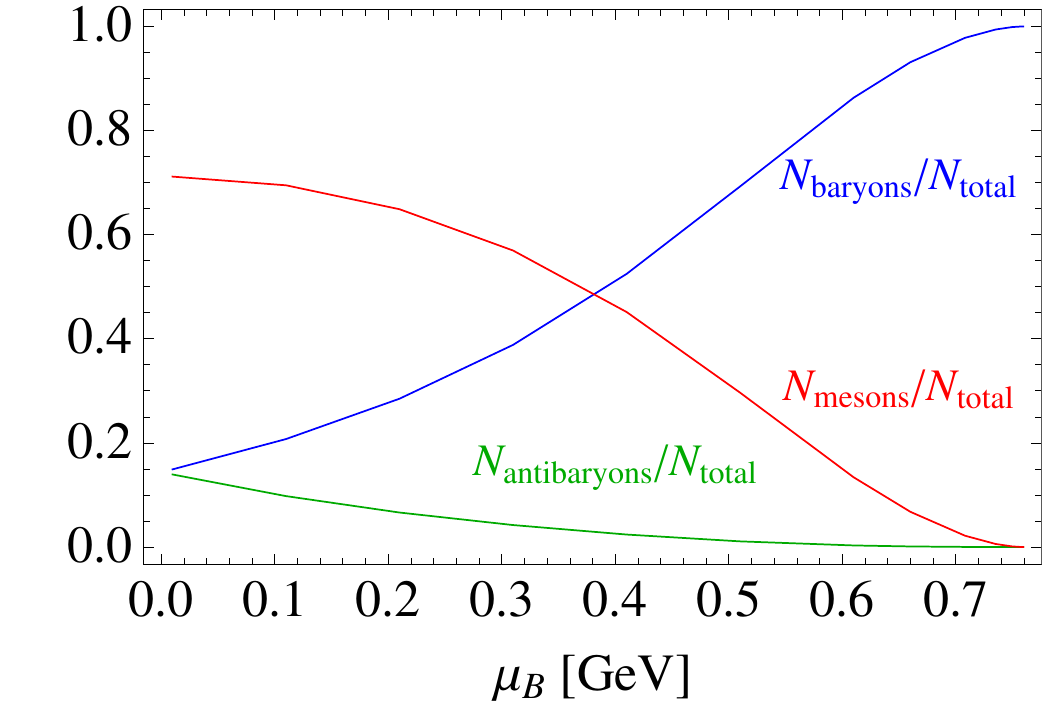} 
     \caption{Same plot as Fig.~\ref{fig:Nbm}, but for the continuous mass spectrum. See caption of Fig.~\ref{fig:Nbm} for further details.}
     \label{fig:Nbm_continuous}
  \end{center}   
\end{figure}

Figure \ref{figTmu_continuous} shows a $T$ vs $\mu$ diagram similar to the one in Figure \ref{figTmu_discretemasses} but for the continuous mass scenario. In this case, the freeze-out line, again 
the solid red line, extends itself in a range of $\mu_B$ similar to the previous case but now reaches a maximum temperature for confinement of $0.06$ GeV at $\mu_B = 0.85$ GeV. Here, the proton 
can be consistently found at somewhat lower chemical potential, around $\mu_B = 0.75$ GeV, but in wider range of temperature, reaching $T=0.05$ GeV before phase-transition occurs, twice the 
transition temperature obtained in the discrete mass approach. 

Figure \ref{fig:Nbm_continuous} shows essentially the same as Figure \ref{fig:Nbm}. Considering that in this scenario the proton is found at $\mu_B = 0.75$ GeV, the model states in Figure 
\ref{fig:Nbm_continuous} that the content of the bag is solely baryonic which means that the model remains consistent.

\section{Effects of Non-extensivity}

\begin{figure}[!ht]
  \begin{center}
     \subfigure[]{\label{figvaryingq_continuum_masses:energyDens}
     \includegraphics[width=0.4\textwidth]{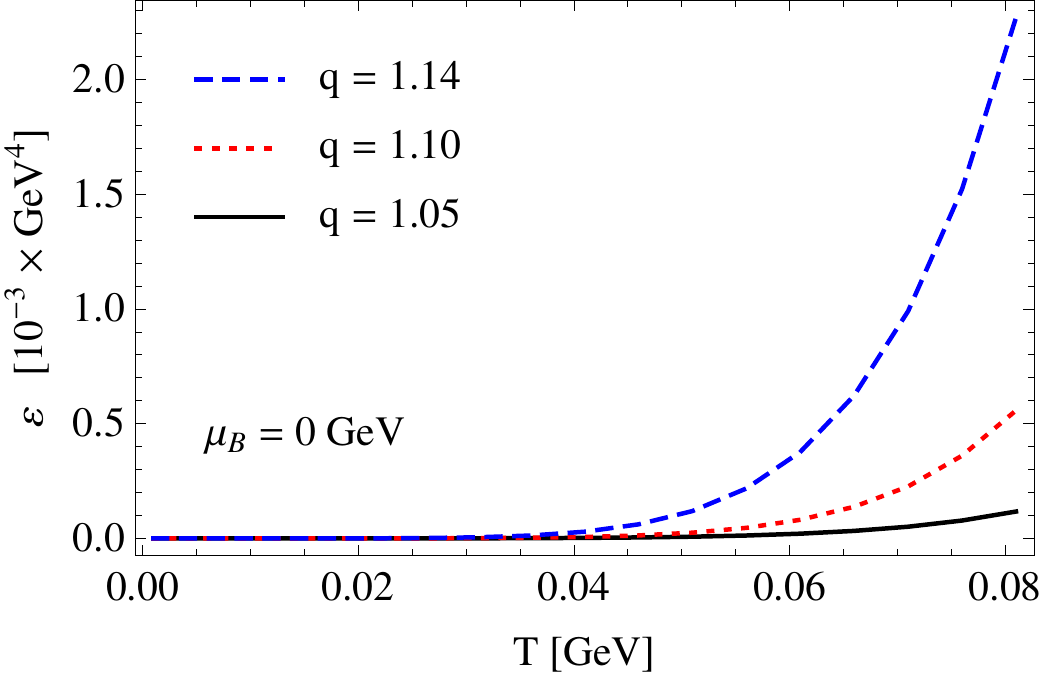} 
     }
     \subfigure[]{\label{figvaryingq_continuum_masses:pressDens}
     \includegraphics[width=0.4\textwidth]{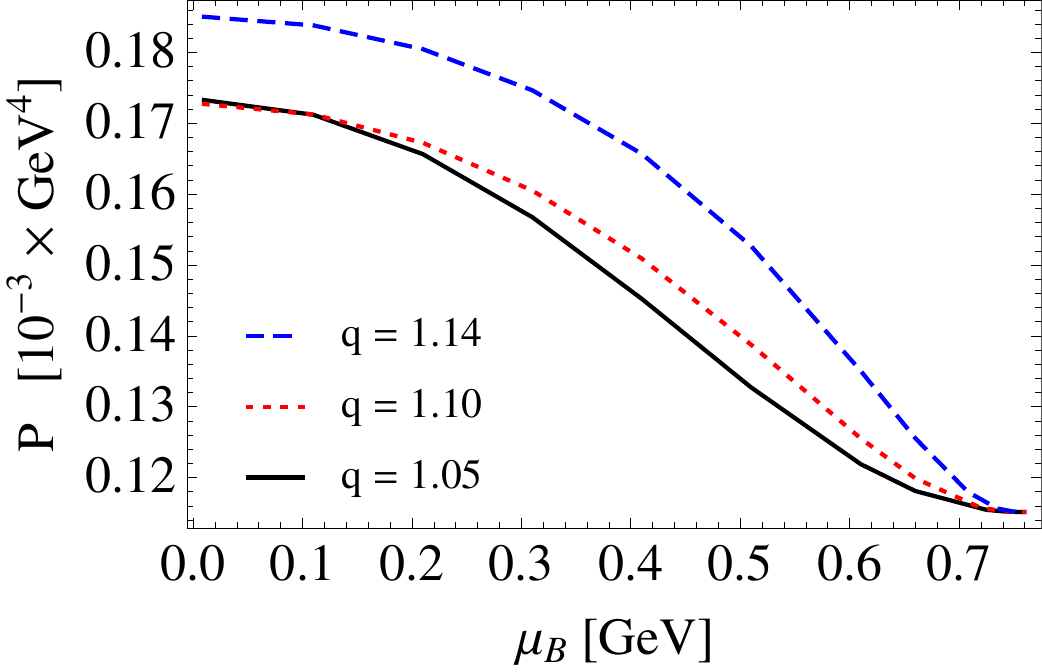}
     } 
     \caption{(a) Energy density as a function of temperature for different values of entropic index, $q$, in the case $\mu=0$. 
     (b) Pressure as a function of chemical potential for different values of $q$ in the case where the temperature $T$ is chosen to 
     keep the total energy fixed to the value $E=m_p$. We consider the case of continuum masses.}
     \label{figvaryingq_continuum_masses}
  \end{center}   
\end{figure}

\begin{figure}[!ht]
  \begin{center}
     \subfigure[]{\label{figvaryingq_discrete_massess:energyDens}
     \includegraphics[width=0.4\textwidth]{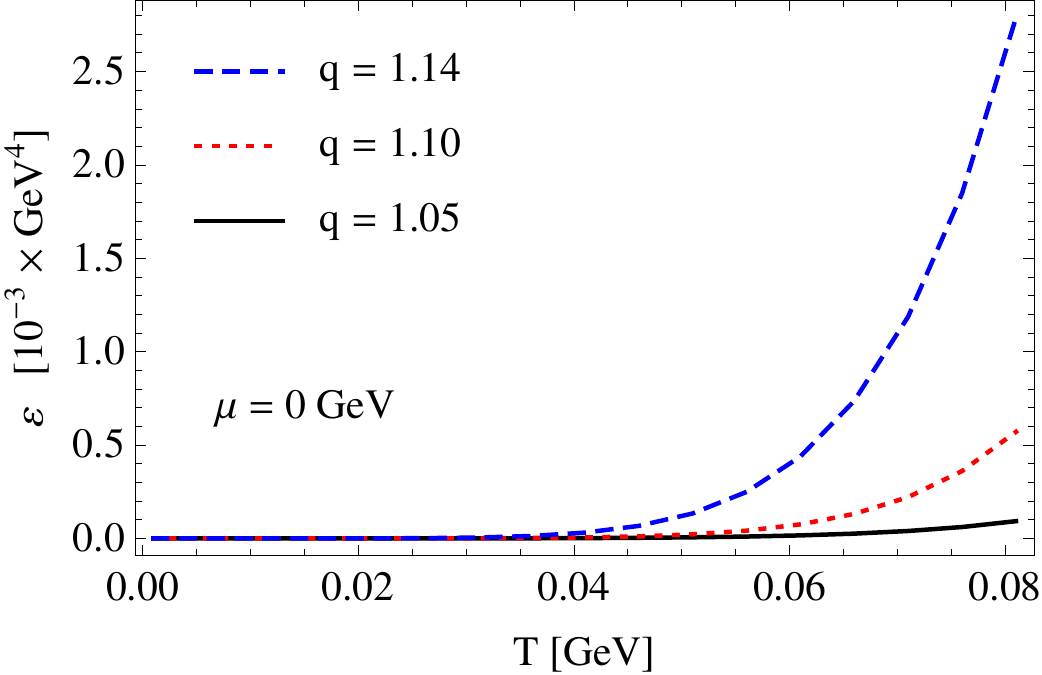} 
     }
     \subfigure[]{\label{figvaryingq_discrete_massess:pressDens}
     \includegraphics[width=0.4\textwidth]{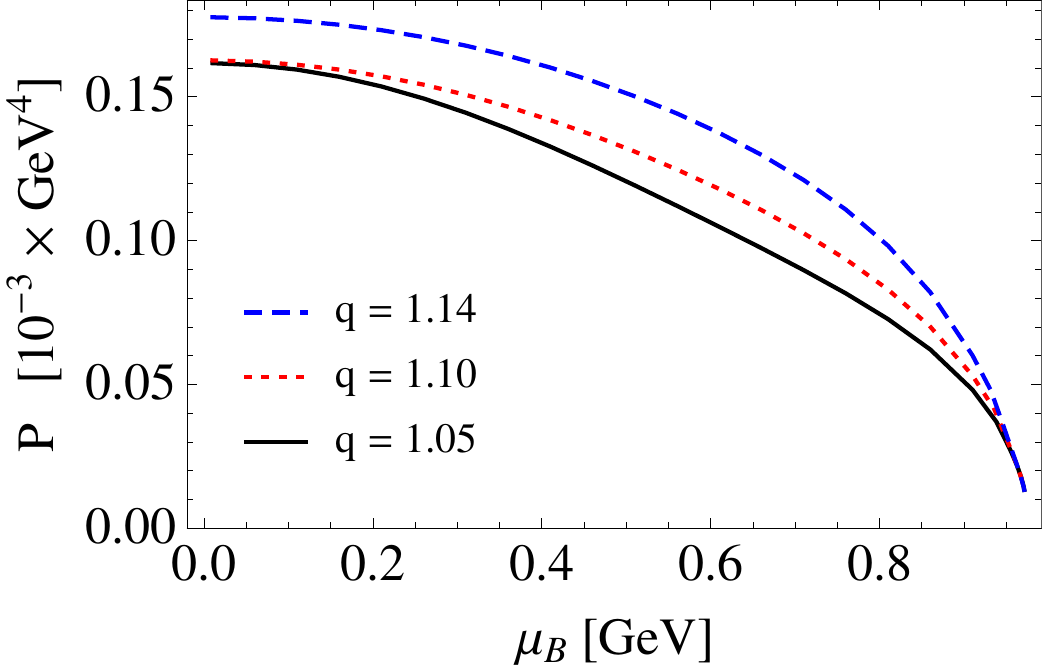}
     } 
     \caption{(a) Energy density as a function of temperature for different values of entropic index, $q$, in the case $\mu=0$. 
     (b) Pressure as a function of chemical potential for different values of $q$ in the case where the temperature $T$ is chosen to 
     keep the total energy fixed to the value $E=m_p$. We consider the case of discrete masses.}
     \label{figvaryingq_discrete_massess}
  \end{center}   
\end{figure}

In order to evaluate the effects of non extensivity in the results obtained here, we present in Figs.~\ref{figvaryingq_continuum_masses} and \ref{figvaryingq_discrete_massess} plots of 
some relevant quantities as obtained using different values for the entropic index $q$, which gives a measure of the non extensivity 
of the system. As $q \rightarrow 1$ the system tends to be extensive and to follow Boltzmann statistics instead of Tsallis statistics. 
We observe that the bag energy density increases faster as the temperature increases for a fixed value of the chemical potential for 
systems with higher entropic indexes with respect to those following extensive statistics, as shown in Figs.~\ref{figvaryingq_continuum_masses:energyDens} and \ref{figvaryingq_discrete_massess:energyDens}. 

\begin{figure}[!ht]
  \begin{center}
     \includegraphics[width=0.4\textwidth]{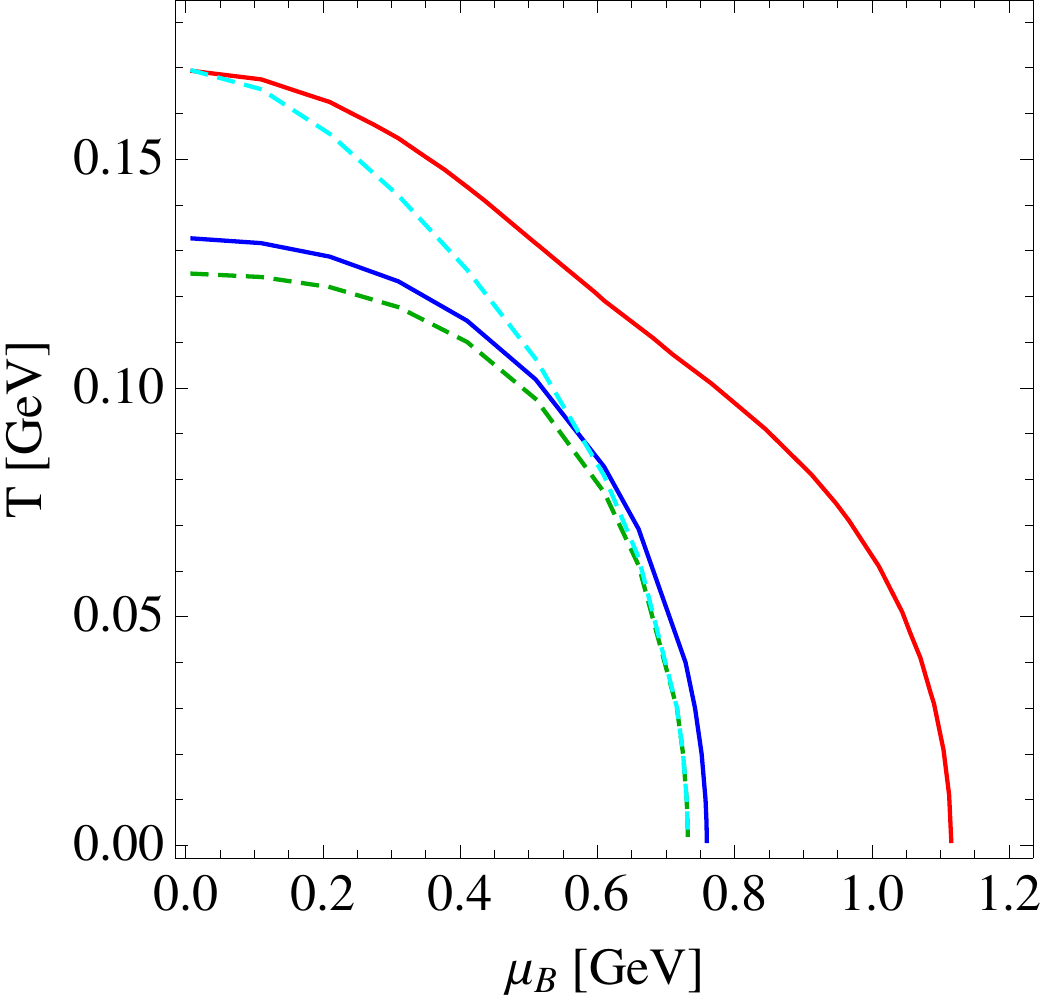}
     \caption{Same plot as Fig.~\ref{figTmu_continuous}, but for Boltzmann-Gibbs statistics $(q=1)$. We have considered the continuous mass spectrum. See caption of Fig.~\ref{figTmu_continuous} for further details.}
     \label{figTmu_BGcontinuous}
  \end{center}   
\end{figure}

\begin{figure}[!ht]
  \begin{center}
     \includegraphics[width=0.45\textwidth]{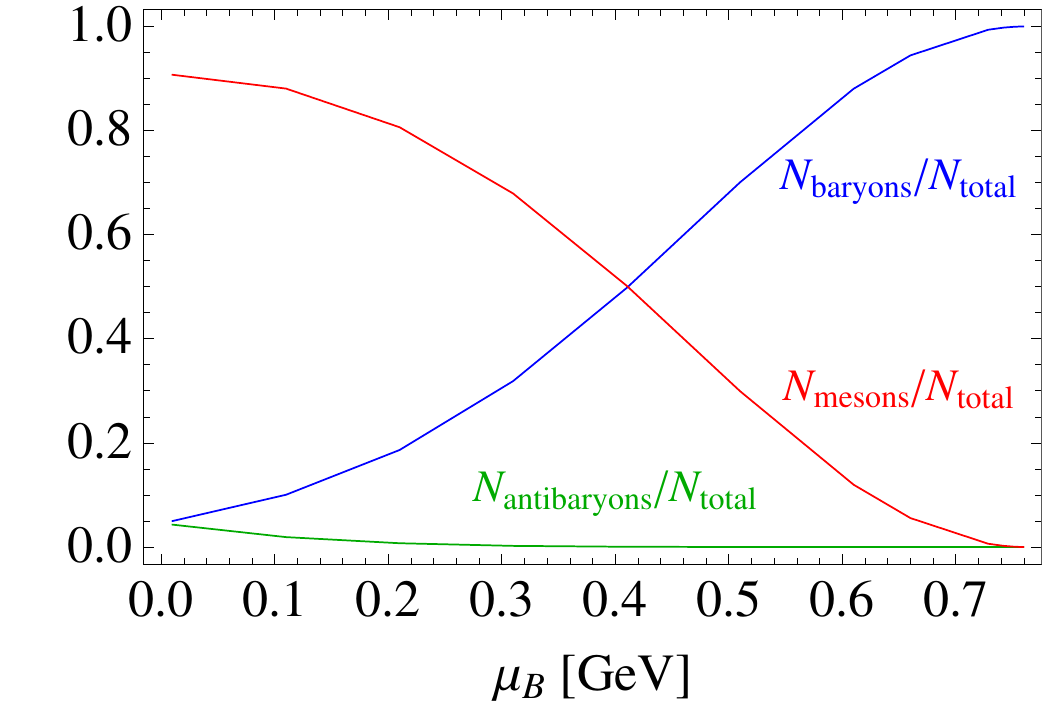} 
     \caption{Same plot as Fig.~\ref{fig:Nbm_continuous}, but for Boltzmann-Gibbs statistics $(q=1)$. We have considered the continuous mass spectrum. See caption of Fig.~\ref{fig:Nbm_continuous} for further details.}
     \label{fig:NbmBGcontinuous}
  \end{center}   
\end{figure}

At this point, it is interesting to investigate what would happen to the system if Boltzmann statistics, instead of Tsallis one, was used. We made the calculations for the continuous and discrete 
mass scenario only since the fixed mass approach has already been shown to be inadequate. The results for continuous mass are displayed in Figs.~\ref{figTmu_BGcontinuous} and~\ref{fig:NbmBGcontinuous}.
We notice a striking difference between the results with Boltzmann-Gibbs statistics in comparison to the results with Tsallis statistics in the T vs $\mu$ diagram, where the existence of the proton 
is possible for any value of T and $\mu$ satisfying the total energy constraint, while in the Tsallis case only a small range of chemical potential would allow the existence of the system in the 
confined region. This results from the fact that the critical temperature is higher in the BG statistics than in the non extensive case. The fractions of fermions and bosons, on the other hand, 
present a similar behavior in both cases.

\begin{figure}[!ht]
  \begin{center}
     \includegraphics[width=0.4\textwidth]{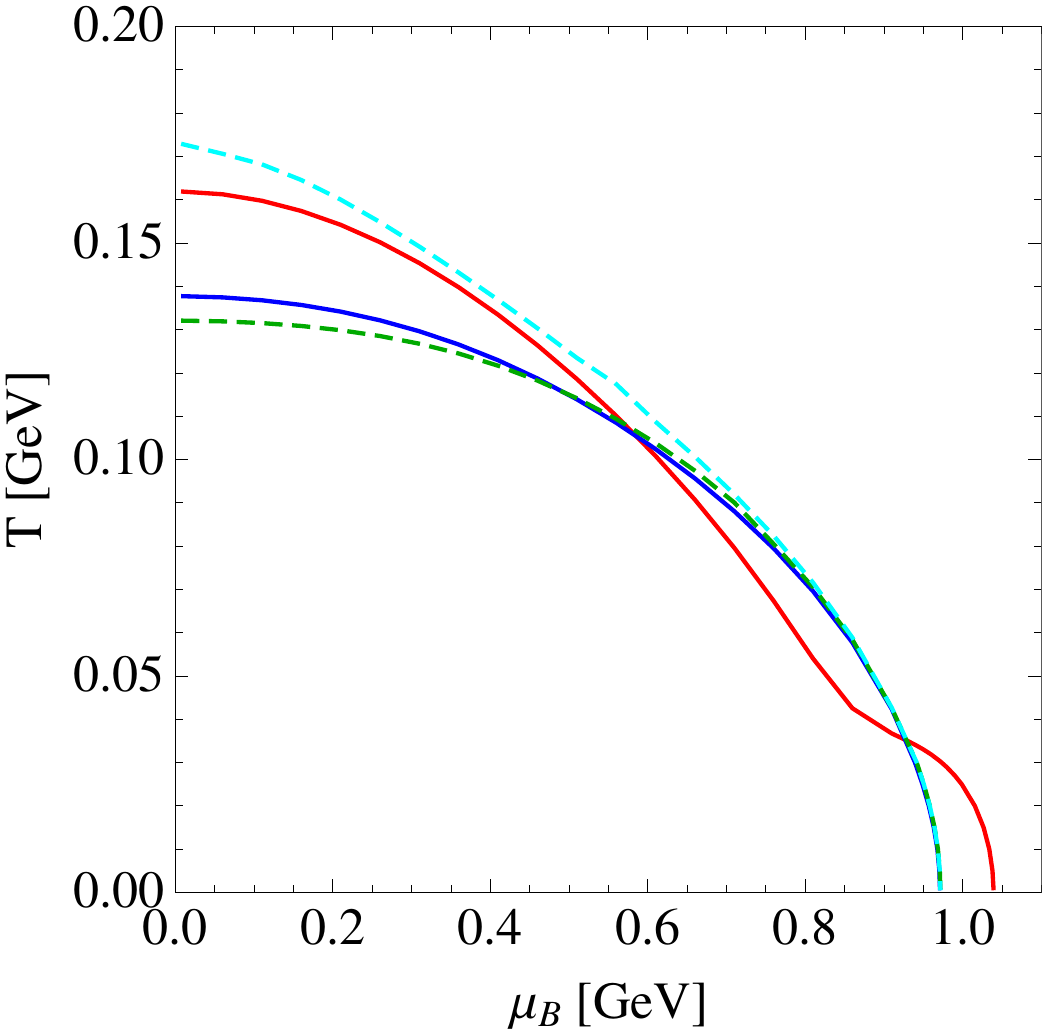}
     \caption{Same plot as Fig.~\ref{figTmu_discretemasses}, but for Boltzmann-Gibbs statistics $(q=1)$. We have considered the discrete mass spectrum. See caption of Fig.~\ref{figTmu_discretemasses} 
     for further details.}
     \label{figTmu_BGdiscretemasses}
  \end{center}   
\end{figure}

\begin{figure}[!ht]
  \begin{center}
     \includegraphics[width=0.45\textwidth]{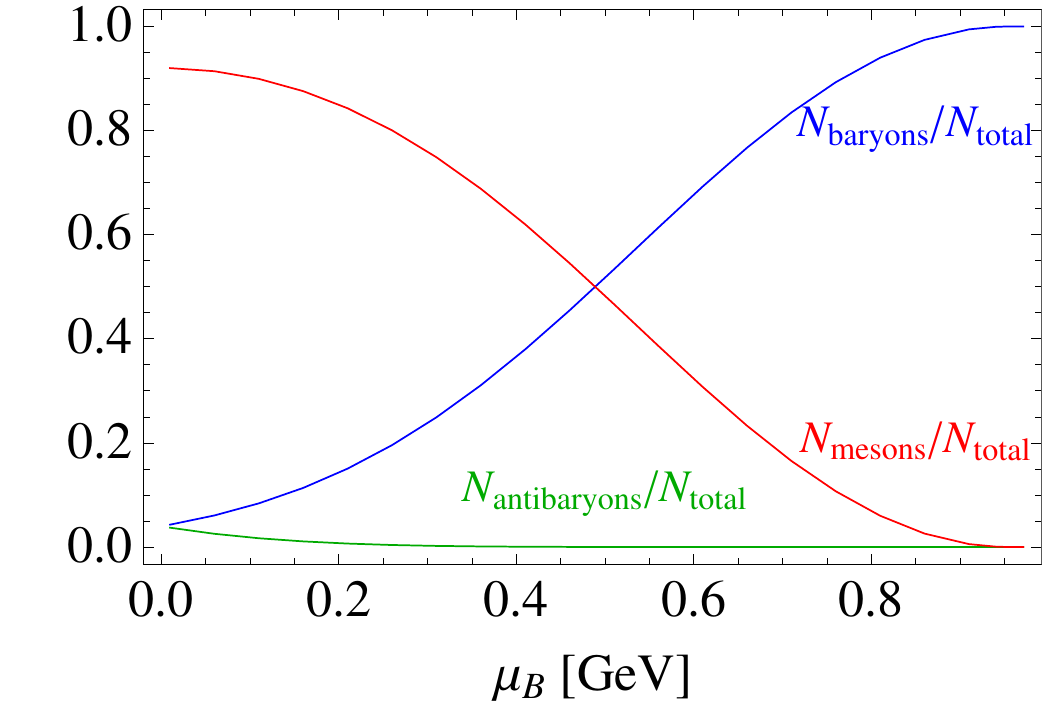} 
     \caption{Same plot as Fig.~\ref{fig:Nbm}, but for Boltzmann-Gibbs statistics $(q=1)$. We have considered the discrete mass spectrum. See caption of Fig.~\ref{fig:Nbm} for further details.}
     \label{fig:NbmBG}
  \end{center}   
\end{figure}

Figures \ref{figTmu_BGdiscretemasses} and \ref{fig:NbmBG} also show the behavior of the T vs $\mu$ diagram and that of the number of particles, respectively, in this case for discrete masses. 
Once more, the number of baryons and mesons exhibit a behaviour similar to the previous cases. Figure \ref{figTmu_BGdiscretemasses}, on the other hand, shows a clearly non-physical situation. 
Following the blue line, for which the total energy is fixed at the proton mass, from lower to higher 
temperature one can see that the system deconfines after crossing the critical line (red line) at 
around $\mu_B = 0.9$ GeV and $T=0.03$ GeV only to confine itself again at $\mu_B = 0.6$ GeV and 
$T = 0.11$ GeV. This reconfinement at a higher temperature is not an observed behaviour of systems undergoing phase transition. 

In summary, Figure \ref{figTmu_BGcontinuous} implies that the proton never undergoes phase transition and Figure \ref{figTmu_BGdiscretemasses} suggests that the proton can de-confine and re-confine 
at a higher temperature. Therefore, these two Figures show that Boltzmann-Gibbs statistics leads to a non-physical conclusion in both scenarios.

\section{Comparison Between Discrete and Continuum Mass Scenario}

\begin{figure}[!ht]
  \begin{center}
     \subfigure[]{
     \label{fig:EnergyDensityContinuous}
     \includegraphics[width=0.45\textwidth]{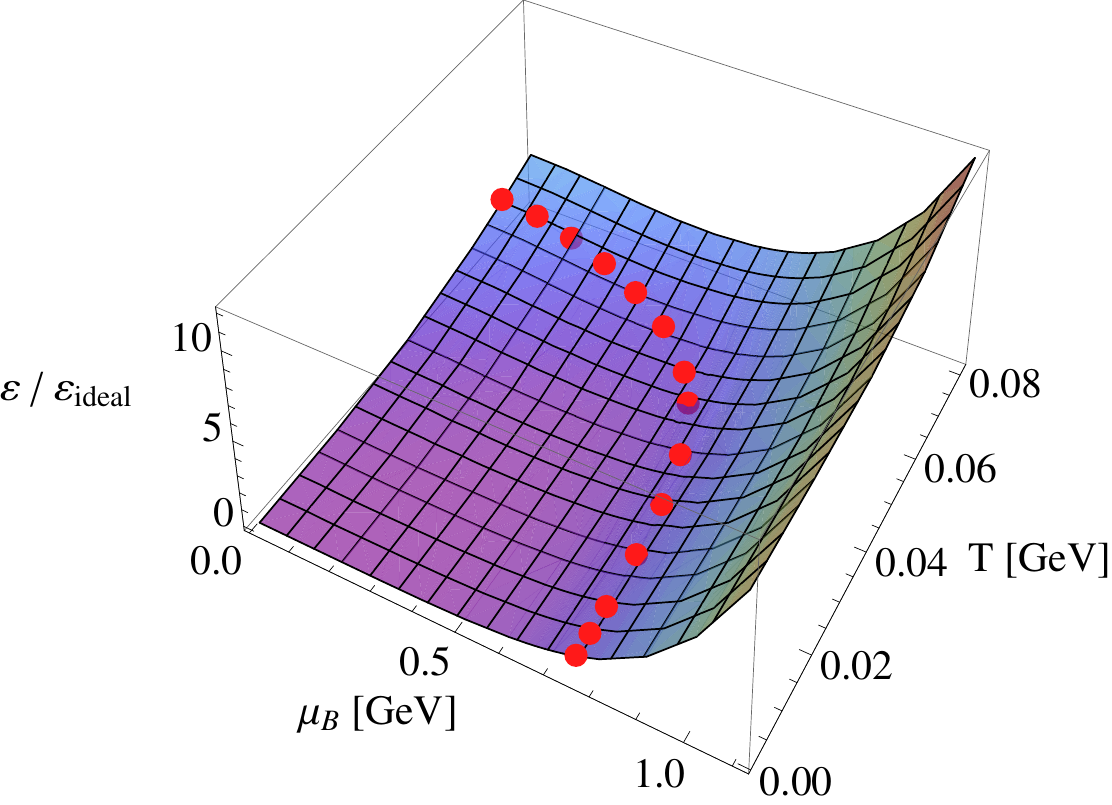} 
     }
     \subfigure[]{
     \label{fig:EnergyDensityDiscrete}
     \includegraphics[width=0.45\textwidth]{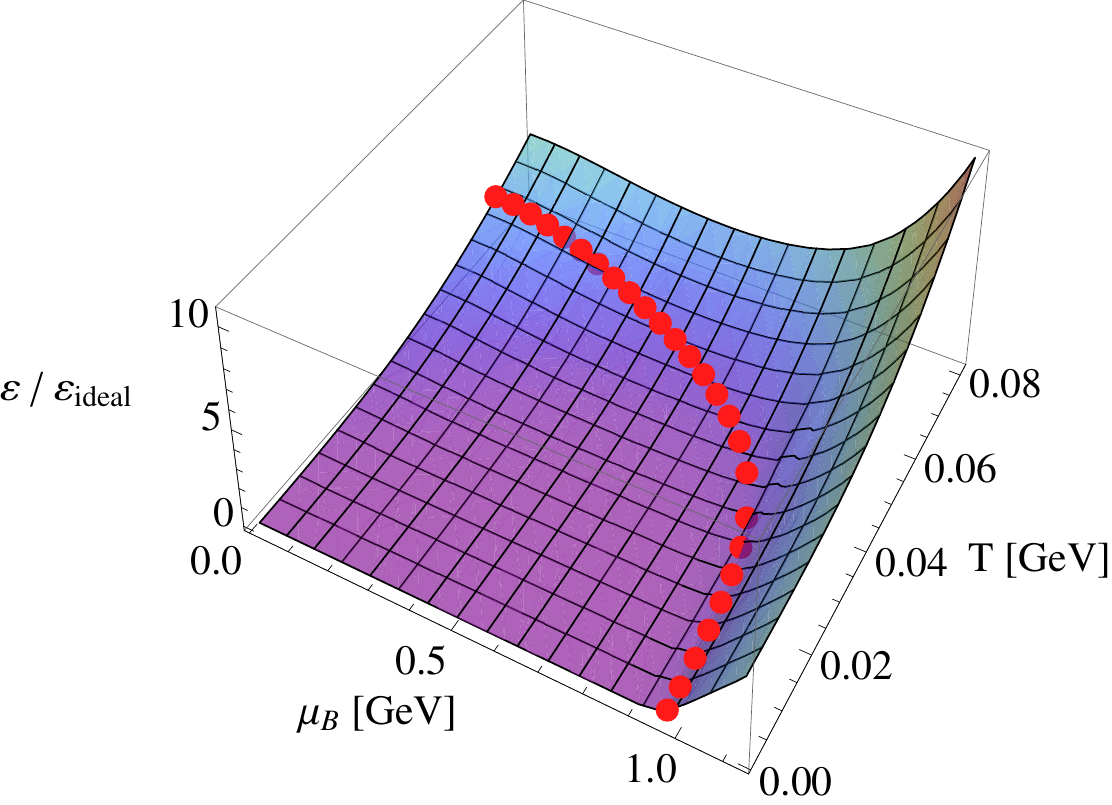}
     } 
     \caption{Energy density (normalized to the ideal gas limit) of a non extensive ideal gas in the case of fixed mass, for $q=1.14$. We show the result for the continuum mass~(a), and discrete mass~(b) spectrum.}
     \label{figE2Dfixedmass}
  \end{center}
\end{figure}

In Fig.~\ref{figE2Dfixedmass} we plot the system 
energy density as a function of $T$ and $\mu$.
As expected, the energy increases with the temperature and with the chemical potential, and the shape reflect the power-law  behavior determined by the q-exponential function.

\begin{figure}[!ht]
   \begin{center}
     \subfigure[]{
     \label{fig:PressureContinuous}
     \includegraphics[width=0.45\textwidth]{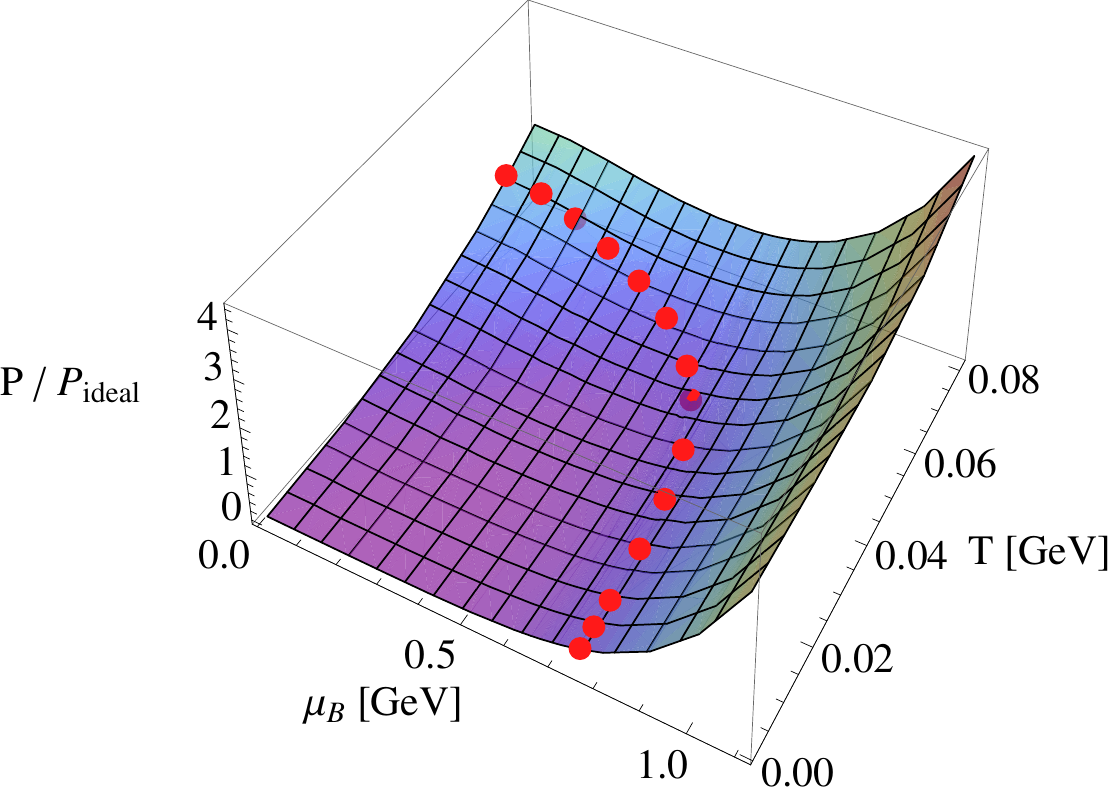} 
     }
     \subfigure[]{
     \label{fig:PressureDiscrete}
     \includegraphics[width=0.45\textwidth]{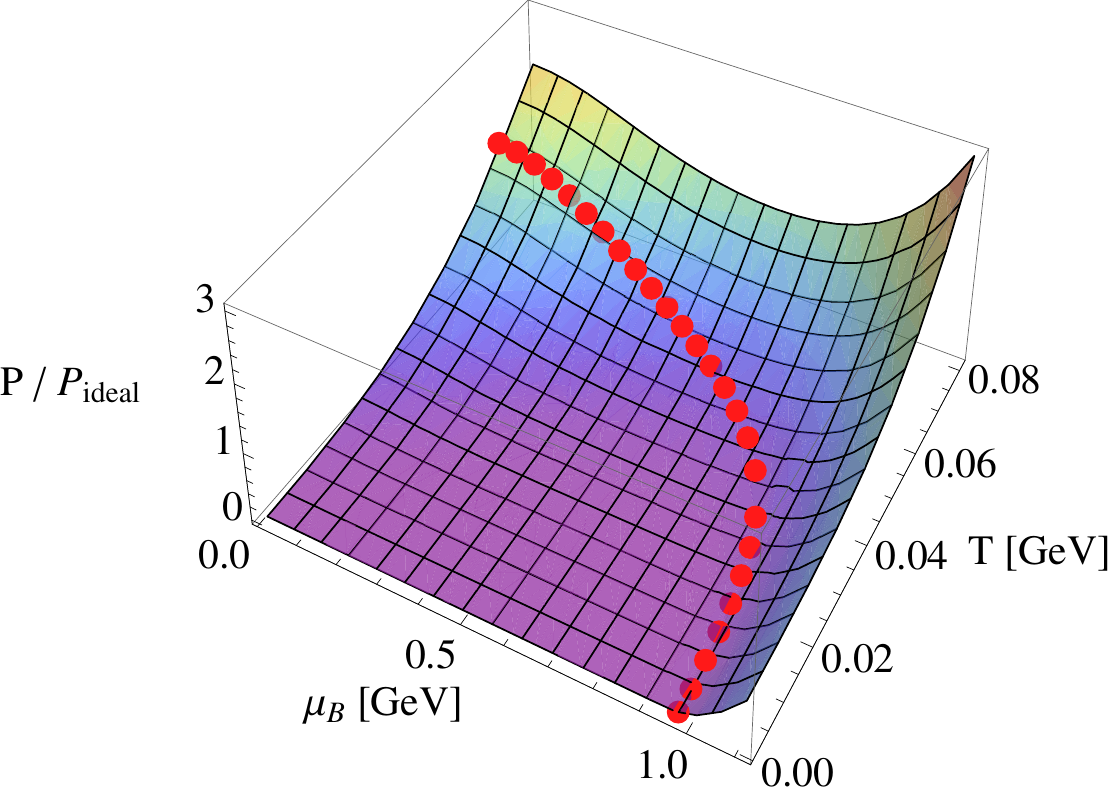}
     }  
      \caption{Pressure (normalized to the ideal gas limit) as a function of temperature and baryonic
        chemical potential in the case of a non extensive gas with
        particles with discrete masses and $q=1.14$. Blue dots
        indicate the region where the gas total energy is equal to the
        proton mass. We show the result for the continuum mass (a) and
        discrete mass (b) spectrum.}
      \label{PressurevsTmudiscrete}
   \end{center}  
\end{figure}

In Fig~\ref{PressurevsTmudiscrete} we show how pressure varies as a function of $T$ and $\mu$, for the case of discrete masses. The plots in Figs.~\ref{figE2Dfixedmass} and \ref{PressurevsTmudiscrete} 
are normalized to the ideal gas limit of the corresponding thermodynamic quantities, which at finite temperature and baryon chemical potential read~\cite{Vuorinen:2003fs}
\begin{equation}
P_{\textrm{ideal}}(T,\mu_B) = \frac{\pi^2}{45} T^4 \left(N_c^2-1 + N_c N_f \left( \frac{7}{4} + 30 \bar{\mu}_B^2 + 60 \bar{\mu}_B^4  \right) \right) \,,
\end{equation}
for the pressure, and $\varepsilon_{\textrm{ideal}} = 3 P_{\textrm{ideal}}$ for the energy density, where $\bar{\mu}_B =  \mu_B/(6\pi T)$. We are considering $N_c = 3$ and $N_f = 3$.

\begin{figure}[!ht]
  \begin{center}
     \includegraphics[width=0.45\textwidth]{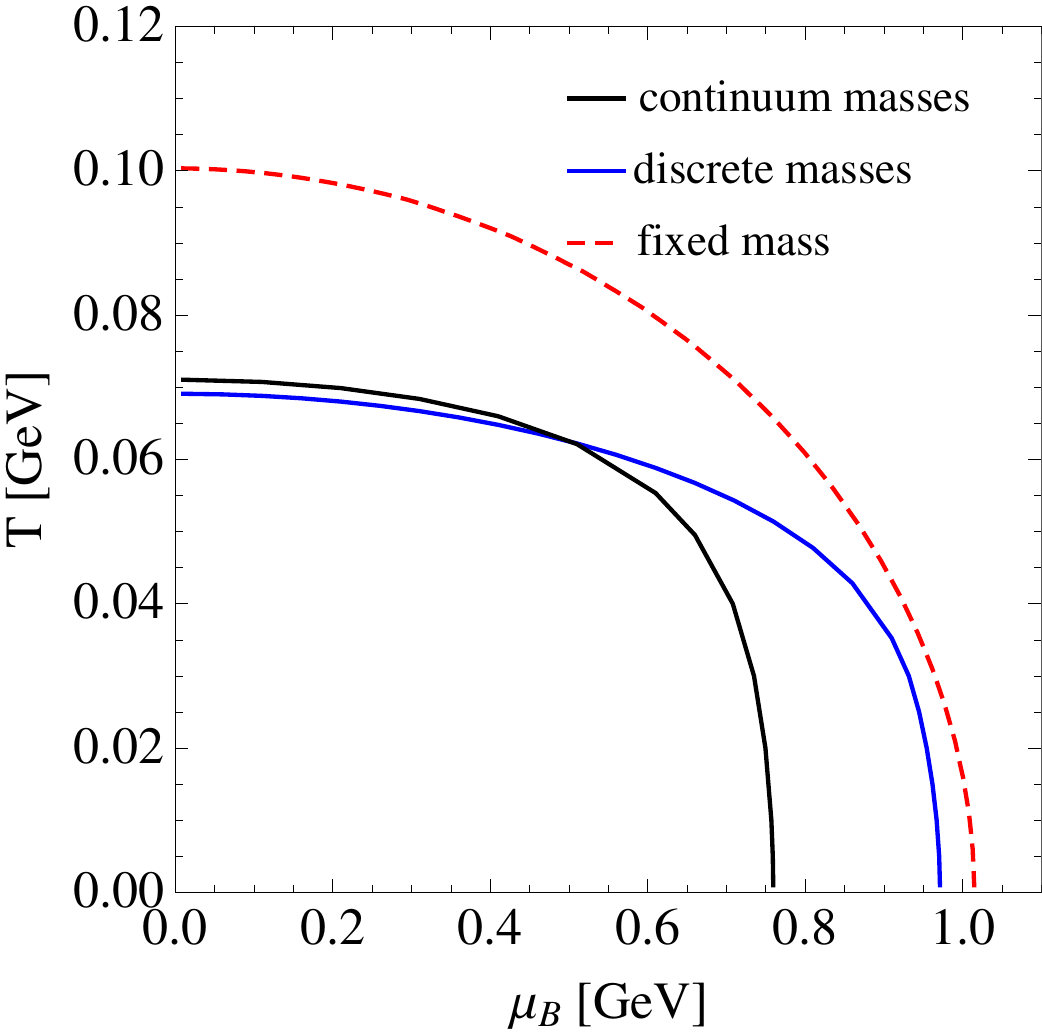}
     \caption{Temperature as a function of baryonic chemical potential
       for a non-extensive gas with total energy inside a volume~$V_p$
       equal to the proton mass and $q=1.14$. For comparison, we
       display the result in three different descriptions: i)
       continuum masses, ii) discrete masses, and iii) fixed mass.}
     \label{fig:Tmu}
  \end{center}   
\end{figure}

As expected, the pressure increases as $T$ and $\mu$ increase. The curve for which $T$ and $\mu$ results in total energy equal
to the proton mass is indicated by blue points. It is possible to observe already in this plot that the region corresponding
to the proton mass is far from the divergent region which shows that the proton structure can be consistently obtained in the model we are adopting.

\begin{figure}[!ht]
  \begin{center}
     \subfigure[]{
     \includegraphics[width=0.4\textwidth]{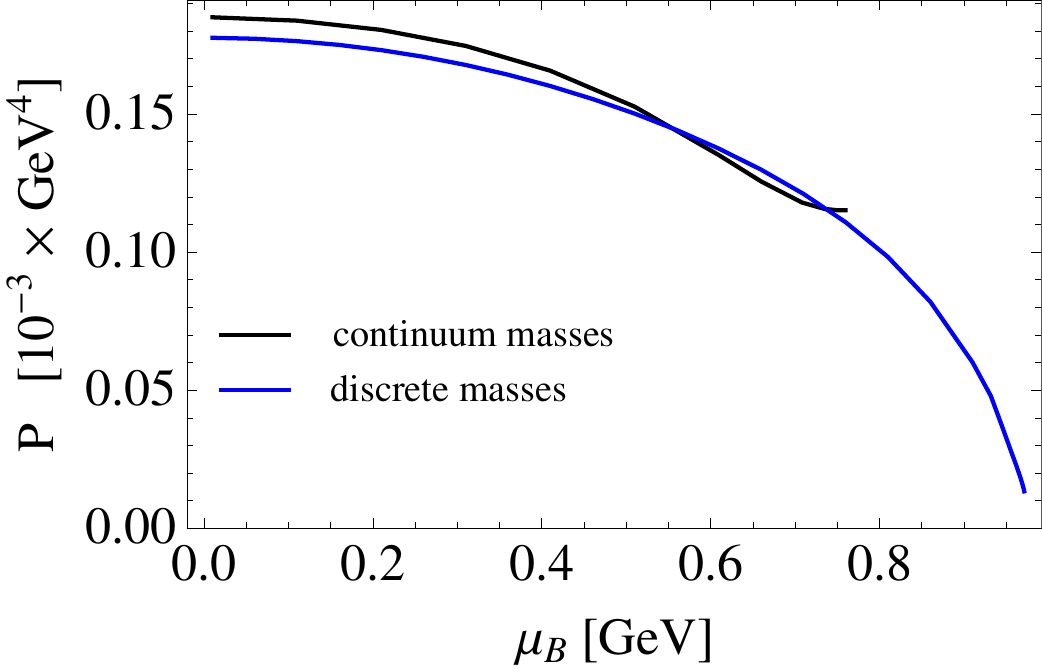}
     }
     \subfigure[]{
     \includegraphics[width=0.4\textwidth]{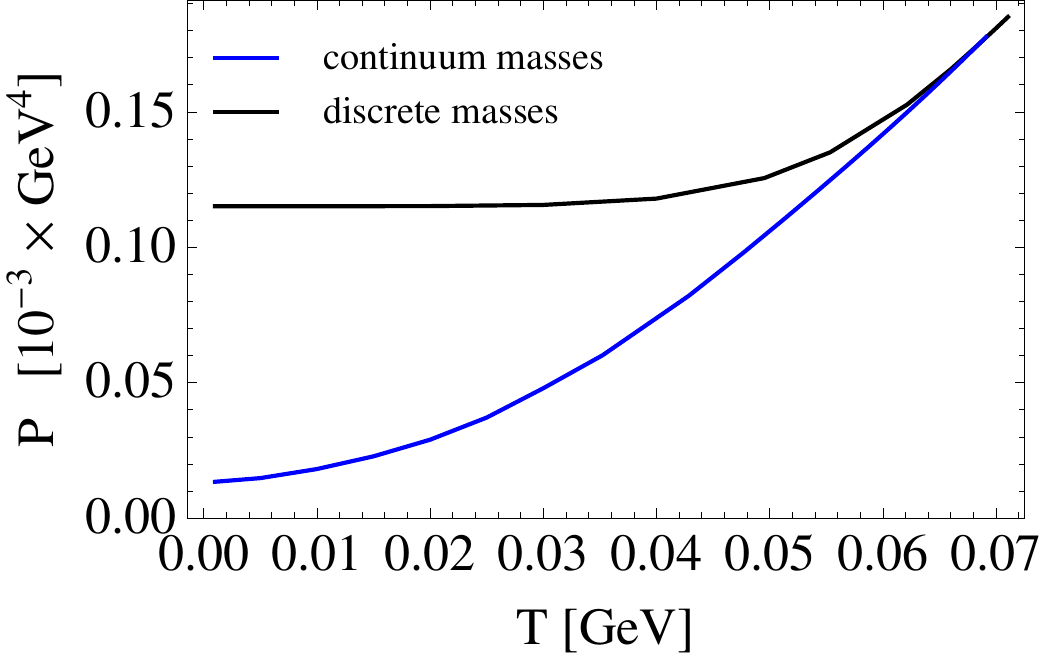}
     } 
     \caption{Pressure as a function of chemical potential (a) and temperature (b) for a non extensive gas with total energy equal to the proton mass and $q=1.14$. We display the result for the
       case of continuum masses (black line) and discrete masses (blue line).}
     \label{fig_PvsmuT_E=mp_discrete_n_continuos}
  \end{center}   
\end{figure}

Fig.~\ref{fig_PvsmuT_E=mp_discrete_n_continuos} shows a comparison between the discrete and the continuous mass approach by showing the pressure as a function of chemical potential and temperature 
for a non extensive gas with total energy equal to the proton mass and $q=1.14$.

\section{Comparison with the MIT-bag model}
\label{subsec:MIT}

Since the MIT Bag Model was modeled based on a fixed mass perspective \cite{MITBag} we now reconsider the fixed mass approach in order to make a useful calculation comparison.

The computation with the fixed mass approach leads to a constant value of $P \simeq 0.332 \cdot 10^{-3} \GeV^4$ for the pressure along the line of $E = m_p$. In fact 
the system seems to behave close to the conformal limit, so that the trace anomaly is vanishing, i.e. $\varepsilon - 3 P \simeq 0$. With these ingredients, 
it is straightforward to compute the bag constant. Using that $E = \varepsilon \cdot V_p = m_p$ together with $P = \varepsilon/3$, one finds
\begin{equation}
P = \frac{m_p}{3 V_p} = (0.135 \GeV)^4 \,.
\end{equation}
This value is consistent with the range of results obtained for $B$ in a previous work \cite{Pedro} and is in good agreement with the value of the bag constant obtained in the context of QCD calculations
\begin{equation}
B = (0.486 \Lambda_{\QCD})^4 = (0.106 \GeV)^4 \,,
\end{equation}
where $\Lambda_{\QCD} = 0.218 \GeV$, see e.g.~Ref.~\cite{Schaden:1998ty}.

\section{Conclusions}

In this work we presented a bag-type model of a hadron assuming fractal structure. By adding the hypothesis that the hadron bag is an ideal gas of strong interacting particles, the non-extensive 
thermodynamics was shown to be the most appropriate theory. Regarding the masses of the particles inside the bag, three scenarios were considered: fixed, discrete and continuous mass, the first 
one being an exercise since no structure was considered for the particles. The discrete and continuous mass scenarios fully applied the hypothesis above. In all situations, pressure, energy and 
number of particles were calculated along with the temperature vs chemical potential diagram in order to thoroughly examine the model.

The consistency of this bag-type model with fractal structure was confronted with a bag containing total energy equal to the proton mass and same proton volume and also with a system satisfying 
the freeze-out line along which $\varepsilon/n = 1$ GeV. It was shown that the proton can be consistently found in both discrete and continuous mass scenarios as a confined system below the freeze-out 
line and below the phase-transition temperature of 61 MeV.

The effects of non-extensivity on energy and pressure were studied and Boltzmann-Gibbs statistics was also applied. It was shown that Boltzmann-Gibbs statistics leads to non-physical conclusions in 
both discrete and continuous mass situations.

Finally, the value of $(0.135\, {\rm GeV})^4$ calculated for the pressure on the bag surface of the proton in this model is consistent with other results found in the literature, including QCD 
calculations \cite{Pedro,Schaden:1998ty}.

\section*{Acknowledgements}

A.D. would like to thank the University of Granada, where part of this work has been done, for the hospitality and financial support under a grant of the Visiting Scholars Program of the Plan Propio de 
Investigaci\'on of the University of Granada. He also acknowledges the hospitality at Carmen de la Victoria. A.D., D.P.M. and T.N.dS. are supported by the Project INCT-FNA Proc. No. 464898/2014-5. A.D. 
and D.P.M. are also partially supported by Conselho Nacional de Desenvolvimento Cient\'{\i}fico e Tecnol\'ogico CNPq (Brazil) under grants 304244/2018-0 (A.D.) and 301155.2017-8 (D.P.M.). 
The work of E.M. is supported by the Spanish MINEICO and European FEDER funds under Grants FIS2014-59386-P and FIS2017-85053-C2-1-P, by the Junta de Andaluc\'{\i}a under Grant FQM-225, and 
by the Consejer\'{\i}ıa de Conocimiento, Investigaci\'on y Universidad of the Junta de Andaluc\'{\i}a and European Regional Development Fund (ERDF) Grant SOMM17/6105/UGR. The research of 
E.M. is also supported by the Ram\'on y Cajal Program of the Spanish MINEICO under Grant RYC-2016-20678.

\end{document}